\documentclass[preprint,review, 12pt, authoryear, natbibg=true]{elsarticle}

\usepackage{amssymb}

\usepackage{amsthm}
\usepackage{amsmath}

\usepackage[left=1in, right=1in]{geometry}

\usepackage{xcolor}
\usepackage{soul}
\usepackage{lineno}

\journal{Ocean Modelling}

\begin{document}

\begin{frontmatter}



\title{Lagrangian study of dispersion and transport  by submesoscale currents at an upper-ocean front}


\author[a]{V. Verma}
\author[a,b]{S. Sarkar}

\address[a]{Dept. of Mechanical and Aerospace Engineering, University of California, San Diego}
\address[b]{Scripps Institute of Oceanography, University of California, San Diego}
              
\begin{abstract}
The three-dimensional transport pathways, the time scales of vertical transport, and the dispersion characteristics (single-, pair- and multi-particle statistics) of submesoscale currents at an upper-ocean front are investigated using material points (tracer particles) that advect with the local fluid velocity. Coherent submesoscale vortex filaments and eddies which dominate submesoscale (0.1 - 10 km) dynamics are found to play a crucial role which is quantified here. {These coherent structures, i.e., submesoscale vortex filaments and eddies, are generated and sustained through non-linear evolution of baroclinic instability.} The collective motion of particles helps identify common features of transport at the front. It is found that the particles in the central region organize into inclined lobes, each associated with a coherent eddy, with a characteristic circulation. Furthermore, the coherent filaments associated with the heavy- and light-edges of the front transfer edge particles to the lobes. This flux of new particles into the central-region causes the particles circulating in the lobes to adjust, which leads to slumping of the front. {The particle motion in the vertical shows multiple time scales -- a fast time scale with $O(10) \, \rm{m}$ vertical displacement within an hour and a slower near-inertial time scale, comparable to the intrinsic time scale of the growing instability. Typically, a particle exhibits the fast motion while moving through vortex filaments.} The overall slumping process is slower than what one might anticipate from the large magnitude of vertical velocity in the filaments and requires a sustained correlation over time between the lateral and the vertical motion. By tracking clouds of particles, we show that their centers of mass downwell/upwell over 1-2 inertial time periods, after which an adjustment follows with a sub-inertial time scale. The dispersion characteristics of the submesoscale turbulent currents using single- and pair-particle statistics have been investigated. The shape change in clusters of four particles reveals filamentogenesis, i.e.  deformation into thin, needle-like structures, which occurs as a rapid process that is complete within approximately an hour.
\end{abstract}

\begin{keyword}
Submesoscale \sep Turbulence \sep Vertical transport \sep Dispersion \sep Baroclinic instability


\end{keyword}

\end{frontmatter}


\section{Introduction} \label{sec:intro}
{Density fronts, ubiquitous in the upper ocean, are an important source of submesoscale dynamics. The dynamics typically occur at length scales of 0.1 - 10 $\, \rm{km}$ and time scale of O(1) day and are characterized by Rossby number, $Ro \equiv U/fL = O(1)$, where $f$ is the Coriolis parameter, and $U$ and $L$ are characteristic velocity and length scales, respectively \citep{ThomasTM:2008, McWilliams:2016}. The submesoscale dynamics plays a significant role in the restratification of the upper ocean and the vertical transport of tracers such as buoyancy, salinity and carbon from the surface ocean to the interior \citep{BoccalettiFF:2007, ThomasTM:2008, FoxKemperFH:2008, OmandDL:2015}. These processes affect the the upper-ocean structure and impact the interactions between the ocean and the atmosphere, thereby influencing the Earth's climate. The submesoscale dynamics also play a significant role in ocean's biochemical cycle by aiding phytoplankton growth through supply of nutrients from the upper thermocline into the surface layer \citep{Mahadevan:2016}. 

Many of the upper-ocean processes driven by the submesoscale dynamics are possible because of their ability to develop large vertical velocity \citep{MahadevanT:2006}, presumably with spatial and temporal coherence.
 This is in contrast to the small-scale turbulent motions, which are relevant for the local mixing, or the balanced mesoscale motions in which the vertical velocity is orders of magnitude smaller. The lateral transport is believed to be dominated by the mesoscale currents and eddies, but the role of submesoscales can be significant as they can provide interconnections between the mesoscale transport barriers and enhance  horizontal spread \citep{HazaOH:2016}. The submesoscales are also important for predicting the dispersion of buoyant pollutants such as oil \citep{DAsaroSK:2018}. An understanding of the organization of vertical velocity and transport pathways is therefore crucial for understanding the submesoscale upper-ocean transport and dispersion processes.} 

Because of their size and relatively fast dynamics, the submesoscale motions have been difficult to investigate using conventional observational methods: ships surveys and satellite remote sensing. However, recent observations employing innovative techniques have uncovered some interesting features of the submesoscale dynamics. By measuring horizontal velocity synchronously along two-parallel tracks, \cite{ShcherbinaDLKMM:2013} were able to calculate the velocity gradient tensor at $O(1)\,\rm{km}$ in the North Atlantic Mode Water region where there is an active submesoscale. Their observations were consistent with dynamics associated with a  predominance of filaments of $O(f)$ cyclonic vorticity in a soup of relatively weak anti-cyclonic vorticity. {The filament structures with cyclonic vorticity are known to develop through frontogenesis that can occur due to straining of the front by a large scale confluent flow. A front can also undergo frontogenesis through non-linear evolution of baroclinic instability (BI) \citep{HoskinsB:1972, Hoskins:1982}. The initial stages of the frontogenetic development of BI at an atmospheric front has been studied in detail by \citet{Mudrick:1974}.} Recent studies have shown that the interaction of a cold filament in thermal wind balance with boundary layer turbulence can drive secondary circulations in the lateral-vertical plane that is frontogenetic and restratifies the filament within a few hours \citep{McWilliamsGMRS:2015, SullivanM:2018}. The ageostrophic circulation in the case of especially strong fronts can lead to nonlinear bores~\citep{PhamS:2018}. Filament structures with cyclonic vorticity were also observed in the northern Gulf of Mexico in an observational campaign utilizing a large number of satellite-tracked surface drifters \citep{DAsaroSK:2018}. The structures were smaller than $1 \, \rm{km}$ in width, separated dense water mass from the light water mass, and were found to be convergent, attracting surface drifters into a line which then wrapped into a cyclonic eddy. The convergence of water mass implies downwelling, and the measured vertical velocity was as large as $1-2 \, \rm{cm\, s^{-1}}$. In comparison, the typical vertical velocity at a mesoscale front is $O(0.01) \, \rm{cm \, s^{-1}}$ \citep{Rudnick:1996}. 

The evolution of BI in upper-ocean density fronts is an important mechanism for generating submesoscale currents. The problem has been studied extensively using large-scale ocean models \citep{CapetMMS:2008a} and turbulence resolving models \citep{SkyllingstadS:2012, HamlingtonRF:2014,StamperT:2017,VermaPS:2019}. Simulating a density front that is initially in thermal wind balance, \citet{VermaPS:2019} (hereafter VPS19) find that the evolution of BI generates long, thin vortex filaments with cyclonic vorticity and downwelling vertical velocity that roll into coherent submesoscale eddies. These submesoscale filaments and the large vertical velocity inside them are similar to the submesoscale filament-like features observed during the the surface drifter measurements of \citet{DAsaroSK:2018}. VPS19 showed that the coherent structures, i.e., vortex filaments and eddies, provide a 3D organization to the secondary circulation whose velocity field suggests that water is transported laterally and vertically across the front.
Although there are organized  3D structures, the actual paths followed by the fluid parcels over time are not apparent from the  instantaneous velocity field as the dynamics is transient. {Furthermore, the spatial pattern of the velocity field changes when the coherent structures are transported by the mean down-front jet.} A Lagrangian framework is better suited for  a study of material transport by the submesoscale, and is the subject of this paper. A related problem is about the time scale of subduction and restratification of the front. The vertical velocity observed in the filaments can be so large as to produce vertical displacement of $O(1)\, \rm{km}$ in a day if sustained in magnitude and direction. However, the restratification is likely to progresses on the time scale of baroclinic instability which is $O(2\pi/f)$ \citep{Stone:1966}. Here, we show evidence of a slow restratification at near-inertial time scale emerging from relatively fast motions in the filament structures.

Lagrangian drifters and floats have been widely used in the ocean for understanding local flow properties and dynamics (e.g. see review article of \citet{LaCasce:2008}). Single-particle metrics are used for calculating the mean flow and eddy kinetic energy \citep{Richardson:1983, Fratantoni:2001, JakobsenRQSH:2003} and the eddy diffusivities \citep{ZhurbasO:2003} in different parts of the ocean and have been utilized for investigating the local transport of tracers \citep{Davis:1985}. The metric of single-particle dispersion  is also a convenient tool for predicting the spread of a particle  from the point of release by a velocity field which has coherent unsteady currents and turbulence.

Particle-pair statistics are often used to probe the scale dependence of dynamics. In a recent  analysis of the trajectories of surface drifters deployed during the Grand Lagrangian Dynamics (GLAD) experiment in the Gulf of Mexico, \citet{PojeOLHH:2014}  found that  the second-order structure function of the velocity field showed power-law 
 behavior from 200 m to 100 km, including the submesoscale range, suggesting the dominance of  local advective dynamics. \citet{BalwadaLS:2016} applied a Helmholtz decomposition of the second-order structure function computed from the GLAD drifters into divergent and rotational components finding that the divergent component dominated  at scales below  $5 \, \rm{km}$, and also computed the third-order structure function. From their analysis, they inferred   forward 3-D energy cascade below $5 \, \rm{km}$, 2-D enstrophy cascade between 5 to 40 km (the deformation radius), and an inverse energy cascade between 40 - 100 km. { \citet{BeronVerraL:2016} examined pair-separation statistics  in the submesoscale range, using synthetic drifter trajectories from data-assimilated NCOM simulations conducted with 1 km horizontal resolution. They found that  the pair separation grew exponentially in accord with non-local   dynamics.
 They further attributed  the discrepancy of their result with the results from GLAD observational drifter trajectories  to the  strong inertial oscillations experienced by the GLAD drifters   and their  limited number of independent samples  with possibly low statistical significance.} In the ocean, internal gravity waves are likely to further complicate the statistical measure arising due to submesoscale dynamics.
 
Two-particle statistics also reveals the spread of the particles about the center of mass (COM) of a particle cloud \citep{Batchelor:1952}. Multiparticle studies have also been used, mainly to measure flow properties such as relative vorticity and horizontal divergence \citep{MolinariK:1975}. Multiparticle statistics using groups of four particles (tetrads) can be used to investigate the changes in the shape of the particle clusters, which result from  straining by the large-scale flows and dispersion by the finescale motions, as shown by \citet{PumirSC:2000}. 
 

In this study, we employ the model front of VPS19 
to investigate the transport and dispersion characteristics of the submesoscale currents, including finescale turbulence,  during the evolution of BI when the dynamics is dominated by the coherent vortex filaments and eddies with localized finescales. VPS19 does not contain strong inertial motions, internal gravity waves or surface forcing, thus providing an ideal setup for studying the dispersion characteristics of submesoscale currents in isolation. Owing to the high resolution (2 m in all three directions) of the simulation, we are able to capture 3D turbulence generated during the evolution of BI. The domain size of 4 km captures a wide range of the submesoscale but not the mesoscale. 
The study is performed in the Lagrangian framework by releasing a large number of tracer particles, which move with the local fluid velocity. 


The paper is structured as follows. In section~\ref{sec:setup}, the numerical method and  the setup of the model front of this work  is described. In section~\ref{sec:sms_structures}, the generation of coherent structures such as vortex filaments and eddies by the baroclinic instability and their organization in 3D are summarized. The details about the tracer-particle simulation are given in section~\ref{sec:prt_setup}. In section~\ref{sec:prt_motions}, the 3D organization of the particles, the typical features of the transport pathways, and the correlation between the lateral and vertical motions of the particles circulating at the front are examined. In section~\ref{sec:vertical_tranport}, the motion of particle clouds, each cloud containing fluid of similar density, is  studied by monitoring the vertical motion of their centers of mass and  the dispersion of the particles about the cloud centers. Additionally, the transport of mean properties such as temperature, subgrid viscosity and kinetic energy associated of with constituent particles are studied. Section~\ref{sec:dispersion} focuses on the dispersion characteristics of the submesoscale currents including the localized   three-dimensional finescale associated with the currents. In this section, single- and pair-particle dispersion statistics are investigated, and shape changes following groups of four particles (tetrads)  are reported. Finally in section~\ref{sec:discusion_conclusion}, conclusions are drawn based on the results, with a brief discussion about their implications.   

\section{Model setup} \label{sec:setup}
The model used here is the same as the one employed by VPS19. Nevertheless, we describe the model for completeness. The model consists of an upper-ocean front in thermal wind balance with a surface jet. The width of the front is $L = 1.2 \, \rm{km}$ and is confined in a surface layer of depth $H = 50 \, \rm{m}$ situated above a strongly stratified thermocline. The density variation is  due to the changes in temperature, and both the quantities are assumed to be related by a linear equation of state, ${\rho}/{\rho_0} = - \alpha T $, where $\alpha = 2 \times 10^{-4} \, \rm{K}^{-1}$ is the coefficient of thermal expansion, $\rho$ is the density deviation from a reference density $\rho_0 = 1028 \, \rm{kg \, m^{-3}}$, and $T$ is the temperature deviation from a reference temperature. The temperature profile is given by  
\begin{align} \label{eq:T}
 T(y, z) =  &-\frac{M_0^2 L}{\alpha g}\left\{1-0.25 \left[1+\tanh\left( \frac{y}{0.5 L}\right) \right] 
           \left[1+\tanh\left(\frac{z+H}{\delta_H}\right) \right] \right\}  \nonumber \\
            &+ \frac{0.5}{\alpha g}\left\{\left( N_M^2+N_T^2 \right)z+ \delta_H \left(N_M^2-N_T^2 \right)  \log\left[\frac{\cosh((z+H)/\delta_H)}{\cosh(H/\delta_H)}\right]\right\}.
\end{align}
Here, the front is assumed to be aligned with the $x$ direction (along-front), and the temperature variation is in the $y$ direction (cross-front) and the $z$ direction (vertical), which also coincides with the axis of rotation; $M_0^2$ is the value of $M^2 = -(g/\rho_0) \partial \rho/ \partial y$ at the center $y = 0$, where $M^2$ is defined analogous to the square of buoyancy frequency associated with the vertical density gradient, $N^2 = -(g/\rho_0) \partial \rho/\partial z$; $N^2_M$ and $N^2_T$ are the squared buoyancy frequencies in the surface layer and the thermocline, respectively; $\delta_H = 5\, \rm{m}$ is a thin region between the surface layer and the thermocline where the temperature profile joins smoothly from its value in the surface layer to that in the thermocline;  $\rm{g}  = 9.81 \, \rm{m}  \, \rm{s}^{-2}$ is the gravitational acceleration. The values of the temperature profile parameters used in the simulation are $M^2_0 = 1.5 \times 10^{-7} \, \rm{s^{-2}}$, $N^2_M = 3.0 \times 10^{-7} \, \rm{s^{-2}}$ and $N^2_T = 10^{-5} \, \rm{s^{-2}}$. 

The surface jet, $U(y,z)$, is constructed from the density field by integrating the thermal wind relation, $\partial U/\partial z = - M^2/f$, where $f = 1.4 \times 10^{-4} \, \rm{s}^{-1}$ is the Coriolis parameter. Additionally, a broadband velocity noise with amplitude of $10^{-4} \, \rm{m} \, \rm{s}^{-1}$ is superimposed to the frontal jet for instigating the instabilities.  

The contours of initial velocity, temperature and potential vorticity over a $y$-$z$ plane are shown in Fig.~\ref{fig:initial_condition}. The potential vorticity is defined as $\Pi = (\boldsymbol{\omega} + f \mathbf{k})\cdot \boldsymbol{\nabla}b$, where $\boldsymbol{\omega}$ is the relative vorticity, $\mathbf{k}$ is a unit vector in the vertical direction and $b = \alpha T g$ is the buoyancy. As shown in Fig.~\ref{fig:initial_condition}, the potential vorticity at the front is initially negative and the setup is unstable to symmetric perturbations.

The evolution of the model front is studied by numerical means, utilizing the  large eddy simulation (LES) approach and solving the non-hydrostatic Navier-Stokes equations under the Boussinesq approximation. Along-front velocity $u_1$, cross-front velocity $u_2$, vertical velocity $u_3$, temperature $T$ and dynamic pressure $p$ are advanced in time $t$ as follows:
\begin{align} \label{eq:NS}
 \frac{\partial u_j}{\partial x_j} &= 0,  \nonumber \\
 \frac{\partial u_i}{\partial t} + \frac{\partial u_i u_j}{\partial x_j} + \epsilon_{ijk} f_j u_k &= -\frac{1}{\rho_0} \frac{\partial p}{\partial x_i} + \alpha T g \delta_{i3} + \nu \frac{\partial^2 u_i}{\partial x_j^2} -  \frac{\partial \tau^{sgs}_{ij}} {\partial x_j},   \nonumber \\
 \frac{\partial T}{\partial t} + \frac{\partial u_j T}{\partial x_j} &= \kappa \frac{\partial^2 T}{\partial x_j^2} - \frac{\partial q^{sgs}_j}{\partial x_j}, 
\end{align}
where $i, \, j, \, k = 1, \, 2, \, 3$,  and a repeated index implies summation; $\nu$ is the molecular viscosity and $\kappa$ is the molecular diffusivity;  $\tau^{sgs}_{ij} = - \nu^{sgs} (\partial u_i/\partial x_j + \partial u_j/\partial x_i)$ is the modeled LES subgrid stress tensor and $q^{sgs}_j = -\kappa^{sgs} (\partial T/\partial x_j)$ is the modeled LES subgrid heat flux, with $\nu^{sgs}$ and $\kappa^{sgs}$ representing the subgrid viscosity and diffusivity, respectively. Parameters $\nu$ and $\kappa$ are related by the Prandtl number  $Pr = \nu/\kappa$; the value of molecular viscosity used is $\nu = 10^{-6} \, \rm{m^2 s^{-1}}$, and the Prandtl number, $Pr = 7$. The subgrid scale $\nu^{sgs}$ and $\kappa^{sgs}$ are related by a turbulent Prandtl number taken to be $Pr^{sgs} = 1$. An alternate notation for the velocity components is also used wherein the along-front, cross-front, and vertical velocity components are expressed as $u$, $v$, and $w$, respectively.

When Eq.~\ref{eq:NS} is scaled by the velocity scale $U_0 = M_0^2 H/f$, the maximum geostrophic jet velocity at the surface, and the buoyancy scale $N^2_M H$, the {\em non-dimensional} parameters are as follows: the Ekman number, $Ek = \nu/fH^2$, the non-dimensional lateral buoyancy gradient, $M_0^2/f^2$, and the Richardson number, $Ri = N^2_M f^2 / M_0^4$. In the present study, $Ri = 0.26$ and $Ek = 2.86 \times 10^{-6}$. The ratio $M_0^2/f^2 = 7.65 $ is comparable to the values used in the studies of \citet{SkyllingstadS:2012} and \citet{HamlingtonRF:2014}. Also, note that the Rossby number, $Ro = U_0/fL$, based on the initial horizontal shear is 0.32 and the Reynolds number, $Re = U_0 H/\nu$, is $2.67 \times 10^{6}$. 

The subgrid fluxes are parametrized following \citet{DucrosCL:1996}. First, the subgrid viscosity, $\nu^{sgs}$, is calculated, and then the subgrid diffusivity of temperature, $\kappa^{sgs}$, is predicted knowing the turbulent Prandtl number, $Pr^{sgs}$. The subgrid viscosity, $\nu^{sgs}$, is computed dynamically at every grid point $(i, j, k)$ using a local velocity structure function $F$:
\begin{equation}
\nu^{sgs} = 0.0014 C_K^{-3/2} \Delta \left[F(x_i, \Delta x_i,t) \right]^{1/2}, \,
\end{equation}
where $C_K = 0.5$ is the Kolmogorov constant, $\Delta = || \Delta x_i ||$ is the magnitude of the filter grid spacing, and
\begin{align}
 F(x,\Delta x_i,t)  = &\frac{1}{4} (||\tilde{\mathbf{u}}_{i+1,j,k} - \tilde{\mathbf{u}}_{i,j,k}||^2 + ||\tilde{\mathbf{u}}_{i-1,j,k} - \tilde{\mathbf{u}}_{i,j,k}||^2 \nonumber \\
                      &+ ||\tilde{\mathbf{u}}_{i,j+1,k} - \tilde{\mathbf{u}}_{i,j,k}||^2 + ||\tilde{\mathbf{u}}_{i,j-1,k} - \tilde{\mathbf{u}}_{i,j,k}||^2). 
\end{align}
For calculating $F(x,\Delta x_i,t)$, the velocity field $\tilde{\mathbf{u}}_{i,j,k}$ is obtained by passing the LES velocity through a discrete high-pass Laplacian filter. The parametrization is efficient in predicting $\nu^{sgs} $, and the values are substantial only at grid points with large velocity fluctuation. Once $\nu^{sgs}$ is known, the subgrid diffusivity $\kappa^{sgs}$ is calculated assuming $Pr^{sgs} = 1$. Note that the dynamic Ducros model used here has been employed in several previous studies, including the oceanic examples of turbulent baroclinic eddies \citep{SkyllingstadS:2012} and the formation of gravity currents from strong fronts \citep{PhamS:2018}. 

The computational domain is also the same. It is a rectangular box bounded by $0 \le x \le 4098 \, \rm{m}$,  $-3073 \, \rm{m} \le y \le 3073 \, \rm{m}$ and $-130 \, \rm{m} \le z \le 0 $. The domain is discretized  in two different ways during during the solution. A uniform grid with $2050 \times 3074 \times 66$ points in employed initially, providing a grid resolution of 2 m in each direction. Later, during the evolution of baroclinic instability, the solution is obtained using a grid which is exactly the same in the horizontal, but has $98$ grid points in the vertical, with uniform stretching such that the grid spacing changes from $0.5 \, \rm{m}$ near the surface to $1.5 \, \rm{m}$ near the bottom of the surface layer. The finer grid resolution in the vertical is needed near the surface to resolve the surface intensified turbulence in the vortex filaments that develops during the nonlinear evolution of BI. As in VPS19, the domain size is chosen to accommodate the growth of the most unstable baroclinic mode \citep{Stone:1966} whose wavelength, $L_b$, and the time scale, $\tau_b$, are:
\begin{equation} \label{eq:BI}
L_b  =  2\pi H \frac{M_0^2}{f^2} \sqrt{\frac{1+Ri}{5/2}}, \quad
\tau_b = \sqrt{\frac{54}{5}} \frac{\sqrt{1+Ri}}{f} \, . 
\end{equation}
With the parameters used in the present study, the chosen domain is large enough to accommodate at least two wavelengths of the most unstable baroclinic mode. 

For obtaining the numerical solution, one also needs to specify the boundary conditions appropriately. We consider the domain to be periodic in the along-front direction. Free-slip on the velocity and no-flux on the temperature are used as the boundary conditions at the  surface ($z=0$) and the lateral boundaries. The bottom boundary conditions are free slip for the velocity and a constant heat flux for the temperature corresponding to the vertical gradient in the thermocline. Sponge layers are employed at the lateral and bottom boundaries to prevent reflection of spurious waves. The sponge layers at the lateral boundaries have a thickness of $64 \, \rm{m}$ and the sponge layer at the bottom boundary is $20 \, \rm{m}$ thick. The governing equations (Eq.~\ref{eq:NS}) are advanced in time using a mixed third-order Runge-Kutta (for advective fluxes) and Crank-Nicolson (for diffusive fluxes). Second-order finite difference discretization is used to compute spatial derivatives. The dynamic pressure is obtained by solving the Poisson equation with a multi-grid iterative method.

\section{Submesoscale structures} \label{sec:sms_structures}
The evolution of the front is discussed in detail in VPS19. Here, we describe the evolution briefly to motivate the Lagrangian studies performed in the remainder of this paper. The front evolves through symmetric and baroclinic instabilities. Initially, the front is unstable to symmetric perturbations as the potential vorticity is negative \citep{Hoskins:1974}. The symmetric instability (SI) grows and forms convection cells nearly aligned with the isopycnals. However, SI does not persist for long. The vertical shear in the convection cells become unstable to secondary Kelvin-Helmholtz instabilities \citep{TaylorF:2009}, which break down into turbulence through tertiary instabilities \citep{AroboneS:2015}. This leads to restratification of the front, making it stable to SI. 

The evolution of the front, to a large extent, is controlled by baroclinic instability (BI), which becomes dominant after SI subsides. The non-linear growth of BI spawns submesoscale coherent structures such as vortex filaments and eddies. Understanding these coherent structures {provides deeper insights into} the dynamics at the front, specifically the vertical and lateral transport, as well as dispersion of tracer particles. Here, we briefly explain how the submesoscale coherent eddies and filaments evolve and how their spatial organization changes in time. The evolution is readily noticeable from the submesoscale flow component obtained  by applying a 2D, low-pass Lanczos filter in $x$- and $y$-coordinate directions, with a cutoff wavenumber, $k_c = 0.04 \, \rm{rad\,m^{-1}}$, and wavelength, $\lambda_c \equiv 2\pi/k_c = 157 \, \rm{m}$. The application of the filter separates the coherent submesoscale from the small scales.

The time evolution of the coherent structures is illustrated in Fig.~\ref{fig:omg3_depth_time}, which shows the submesoscale vertical vorticity at $10 \, \rm{m}$ (Figs.~\ref{fig:omg3_depth_time}a-c) and $30 \,\rm{m}$ (Figs.~\ref{fig:omg3_depth_time}d-f) depth at different times, $t = 57.2 \, \rm{h}, 75 \, \rm{h},$ and $84.9 \, \rm{h}$. The filaments of cyclonic vorticity connected to the heavy edge of the front and wrapping into coherent eddies in the central region can be observed in Fig.~\ref{fig:omg3_depth_time}a plotted at $t = 57.2 \, \rm{h}$. At this time, the vorticity filaments have just begun to roll up, and eddies are small in size, slightly larger than the width of the filaments, $O(100)\,\rm{m}$. The eddies are vertically coherent and can be identified at $30 \, \rm{m}$ depth in Fig.~\ref{fig:omg3_depth_time}d. As the instabilities evolve, the vorticity filaments grow in length and the eddies become larger in diameter (Figs.~\ref{fig:omg3_depth_time}b,c). Moreover, the structures are advected by the mean jet velocity, which is in the negative-$x$ direction in the present model. In Fig.~\ref{fig:omg3_depth_time}a, three small eddies connected to the heavy-edge vortex filaments can be observed. However, Fig.~\ref{fig:omg3_depth_time}b, plotted at $t = 75 \, \rm{h}$, shows two relatively larger eddies, whereas the one situated between them does not grow. This represents a merger of the two cyclonic eddies on the left side of the panel as well as amalgamation of surrounding cyclonic vorticity into the growing vortices. The vortex merger is nearly complete at $t = 84.9 \, \rm{h}$ where the imprint of the eddy, which was initially between the other two, is weak at both $10 \, \rm{m}$ and $30 \, \rm{m}$ depth (Figs.~\ref{fig:omg3_depth_time}c,d). At depth, there are vorticity filaments attached to the light edge of the front that wrap around the eddies. The light-edge filaments can be observed in Figs.~\ref{fig:omg3_depth_time}e,f plotted at $30 \, \rm{m}$ depth, and they wrap around the eddies at the side opposite to where the heavy-edge filaments join with the eddies.

The three-dimensional organization of the coherent structures at a late time ($t = 84.9 \, \rm{h}$) is visualized in Fig.~\ref{fig:coherent_structures_3d}, where the iso-surfaces of Q are plotted. In this figure Q is calculated using the submesoscale velocity fields, i.e., $\tilde{Q} = (\overline{\Omega}_{ij}^2 - \overline{S}_{ij}^2)/2$, and the iso-surfaces are plotted at $\tilde{Q}/f^2 = -0.4$ and $0.4$; here, $\overline{\Omega}_{ij} = (\partial \overline{u}_i/\partial x_j - \partial \overline{u}_j/\partial x_i)/2$ and $\overline{S}_{ij} = (\partial \overline{u}_i/\partial x_j + \partial \overline{u}_j/\partial x_i)/2$ are the second-order rotation and strain-rate tensors, respectively. Thus, regions with positive $\tilde{Q}$ represent rotation-dominated flow at the front, whereas those with negative $\tilde{Q}$ represent strain-dominated flow. The vertical coherence of the submesoscale coherent eddies is evident from the figure, as these structures dominated by cyclonic vorticity appear columnar. The vortex filaments connected to the heavy edge (on the negative-$y$ side) of the front can also be observed. The filaments are shallow at the end connected to the heavy edge; however, they grow deeper as one moves along their length towards the core of the eddies. Surrounding the eddies, the filaments appear to span the entire depth of the front. The filaments have regions which are dominated by strain as well as those where rotation dominates. In the neighborhood of the eddies, there are strain- and rotation-dominated vertical layers which are arranged alternately. We also note that the vortex filaments connected to the light-edge of the front are not obvious in the Q-visualization, possibly due to weaker cyclonic vorticity ($\sim f$) at depth.     

\section{Setup of particle tracking} \label{sec:prt_setup}
In this paper, the main focus is to understand  dispersion and  transport by the submesoscale currents generated by BI. Such a focus necessitates studies of an essentially Lagrangian nature which require the knowledge of how  material points, i.e.,  the tracer particles, move under the influence of submesoscale currents. To this end, tracer particles were introduced in the flow at $t = 57.2 \, \rm{h}$. At this time, the vortex filaments have formed at the front and have begun to wrap into eddies (Figs.~\ref{fig:omg3_depth_time}a,d). The particles are placed at the nodes of a regular lattice over a rectangular subdomain that occupies the entire domain length in the along-front ($x$ direction), $-1.6 \, \rm{km} \le y \le 1.6 \, \rm{km}$ in the lateral ($y$ direction) and $-70\, \rm{m} \le z \le -2 \, \rm{m}$ in the vertical (z direction) with resolution $16 \, \rm{m} \times 16 \, \rm{m} \times 2 \, \rm{m}$. For multi-particle analysis, additional particles are released at $10 \, \rm{m}$ and $30 \, \rm{m}$ depth. Before introducing the particles, the simulation run with a uniform vertical grid of resolution $2\, \rm{m}$ is interpolated to the higher vertical resolution grid  (0.5 m near the surface to 1.5 m near the bottom of the upper-ocean front) at $t \approx 56 \, \rm{h}$. Tracer particles are seeded in the domain at $57.2 \, \rm{h}$ at which point the simulation on the grid with high vertical resolution has progressed by  about an hour. 

The tracer particles are passive and, by definition,  move with the local fluid velocity. Thus, the position of a tracer particle $\mathbf{x}_p = (x_p, y_p, z_p)$ is expressed as
\begin{align} \label{eq:prt_eqn}
 \frac{d \mathbf{x_p}}{dt} = \mathbf{u}_f(\mathbf{x}_p, t),
\end{align}
where $\mathbf{u}_f(\mathbf{x}_p, t)$ is the fluid velocity at the particle's position. The trajectories of the particles are computed by integrating Eq.~\ref{eq:prt_eqn}. The time integration is performed following a third-order Runge-Kutta (RK3) scheme, and the particle velocity $\mathbf{u}_f(\mathbf{x}_p, t)$ is obtained by the fourth-order Lagrange interpolation of a cell-centered velocity field. The Navier-Stokes solver stores the velocity components at the edge centers, and the cell-centered velocity is obtained by the linear interpolation. A CFL value smaller than one is ensured for the particle advection to get a stable numerical trajectory.

\section{{Advection of tracer particles}} \label{sec:prt_motions}
The trajectories followed by individual tracer particles moving in a time-varying velocity field at the front are complex and can differ considerably even for particles released close to each other. Nevertheless, the particles are strongly influenced by the coherent structures, which leads to an overall organization in their motions as elaborated below. Anticipating differences in the transport of particles in the frontal zone and at the edges, it will be convenient at times to distinguish among particle groups according to their cross-front ($y$) locations as follows: (i) central-region particles  released in $-500 \, \rm{m} \le y \le 500\,\rm{m}$, (ii) heavy-edge particles released in $y < -500 \, \rm{m}$ and (iii) light-edge particles released in $y > 500 \, \rm{m}$.  

The influence of coherent vortex filaments and eddies on the transport and organization of tracer particles at the front is illustrated in Figs.~\ref{fig:transport_structure_3D} and ~\ref{fig:downwelling_upwelling}, using particles released at 10 m and 40 m depth. Figure~\ref{fig:transport_structure_3D} shows temperature, lateral velocity, and vertical velocity of the particles at $t = 84.9 \, \rm{h}$, which is $\sim 28 \, \rm{h}$ after they were released. An examination of the particles reveals that they are organized into two lobes (LB1 and LB2) within the front, each associated with a coherent eddy. As the eddies move along the front, so do the lobes. We also notice that the lobes are stratified (Figs.~\ref{fig:transport_structure_3D}a,b), with warm particles constituting the upper portions of the lobes, facing the lighter side of the front, and the cold particles constituting the undersides. Recall that the along-front velocity is vertically sheared, and the near-surface particles in the lobes have larger magnitude of along-front velocity ($u_p$)  compared to those near the bottom, whose velocity magnitude is nearly zero. The along-front motions of the lobes therefore leads to a complex circulation of the constituent particles. The particles appear to circulate clockwise in the inclined lobes when viewed from the top, having negatively correlated lateral and vertical motions. This circulation has been illustrated by plotting lateral ($v_p$) and vertical ($w_p$) velocities of the particles: Figs.~\ref{fig:transport_structure_3D}c,e for the particles released at $10 \, \rm{m}$ depth and Figs.~\ref{fig:transport_structure_3D}d,f for the particles released at $40 \, \rm{m}$ depth. Negatively correlated $v_p$ and $w_p$ of the lobe particles can be observed in these figures. For example, particles at the front side LBF (marked by circles with dots in Fig.~\ref{fig:transport_structure_3D}c) of the lobe LB2 have positive $v_p$. At the same time, the overall vertical motion in LBF is downwelling, as $w_p$ is mostly negative (Figs.~\ref{fig:transport_structure_3D}e,f). These particles moving along the lobe surface in the positive-$y$ direction move downwards. Note that the back-to-front direction is down-front, i.e. oriented along the jet velocity, which is the negative-$x$ direction in the present configuration. Eventually, the lateral velocities become negative, when the particles reach to the back side LBB (marked by circles with crosses in Fig.~\ref{fig:transport_structure_3D}c) of the lobe LB2. The overall vertical velocity becomes positive in LBB, as seen in Figs.~\ref{fig:transport_structure_3D}e, and the particles climb up the lobe. In general, the particles circulate within the same lobe, especially when the interaction between the neighboring eddies is weak.


{It is worth noting that the correlation between $v_p$ and $w_p$ is such that the associated circulation in the lateral-vertical plane follows the isopycnal slope. In the present case, the lateral density gradient points in the negative-$x$ direction and, therefore, $v_p$ and $w_p$ have negative correlation.

The coherent filaments transfer edge particles to the lobes. There are two light-edge filaments (LEF1 and LEF2) that can be identified in Fig.~\ref{fig:transport_structure_3D}b. The filaments have mostly positive $w_p$ (Fig.~\ref{fig:transport_structure_3D}f) and lift the warm-edge particles towards the surface and transfer them to the upper layers of the stratified lobes. Although not visible in the plots shown in Fig.~\ref{fig:transport_structure_3D}, there are coherent filaments connected to the heavy edge of the front, as well. The role of the filament structures on the transport is further examined in Fig.~\ref{fig:downwelling_upwelling} by plotting the heavy-edge particles released  at $10 \, \rm{m}$ depth (Fig.~\ref{fig:downwelling_upwelling}a) and the light-edge particles released at $40 \, \rm{m}$ depth (Fig.~\ref{fig:downwelling_upwelling}b) at $t = 84.9 \, \rm{h}$, after a flight time of $ \sim 28 \, \rm{h}$. In Fig.~\ref{fig:downwelling_upwelling}a, the effect of two coherent filaments connected to the heavy edge of the front is apparent as downwelling of the heavy-edge particles (HEF1 and HEF2) to the underside of the lobes. Once within the lobes, the particles undergo motions that are characteristic of the lobe as described previously in this section. The light-edge filaments LEF1 and LEF2, also marked in Fig.~\ref{fig:transport_structure_3D}b, can be observed in Fig.~\ref{fig:downwelling_upwelling}b. Comparing Figs.~\ref{fig:downwelling_upwelling}a,b shows that the light-edge filaments are located between the heavy-edge filaments. Additionally, we find that the particles moving through the light-edge filaments loop back to the bottom of the front after they are lifted upwards. The explanation is that the light-edge filaments transfer the particles to the lobe, upon which they subduct through the downwelling limb of the lobe. Interestingly, some particles in LEF2 detach from the main branch near the surface and spread laterally under the influence of the near-surface circulation. The particles detaching from the main branch near the surface are enclosed within the rectangular box shown in Fig.~\ref{fig:downwelling_upwelling}b. 

The transport processes mediated through filaments and eddies and the circulation of fluid particles organized into lobes, as discussed in Figs.~\ref{fig:transport_structure_3D} and Fig.~\ref{fig:downwelling_upwelling}, are further illustrated by the schematic shown in Fig.~\ref{fig:transport_schematic}. The schematic depicts the overall motions through the heavy- and light-edge filaments, which transfer fluid particles from the edges to the lobe structures in the central region of the front; the circulations within the lobes are also shown.

The flux of edge particles into the central region causes the local particles to adjust, leading to the restratification of the front. In order to understand the subsequent adjustment of the front, it is key to characterize how the particles that were released in the central region redistribute spatially as time progresses. Figure~\ref{fig:front_adjustment} shows the time evolution of the particles released in the central region, with $-500 \, \rm{m} < y < 500 \, \rm{m}$ and $z > -50 \, \rm{m}$. For identifying their distribution, the particles are sampled in the cells of a rectangular grid that has lateral resolution $\Delta_s y = 16 \, \rm{m}$ and vertical resolution $\Delta_s z = 2 \, \rm{m}$. The distributions are plotted at three different times: $t = 69 \, \rm{h}, 84.9 \, \rm{h}$, and $99 \, \rm{h}$, about $12 \, \rm{h}, 28 \, \rm{h}$ and $40 \, \rm{h}$ after their release. As the front evolves through BI, the particles contained in the region $-500 \, \rm{m} < y < 500 \, \rm{m}$ and $z > -50 \, \rm{m}$ spread laterally while remaining confined in the central region, indicating slumping of the front. The number of particles in the cells within the particle cloud away from the edges does not change considerably with time, consistent with the incompressibility of the flow. Thus, the edge particles that are brought into the front are primarily organized in regions above or below the central-region particles. The isotherms corresponding to the mean temperature of the sampled particles are also plotted. They reveal that the stratification is maintained during the lateral spread of the particles. Further, the stratification becomes stronger with increasing time.



The aforementioned features of the Lagrangian transport can be identified in individual particle trajectories. In Fig.~\ref{fig:prt_trajectories}, the trajectories $x_p(t)$, $y_p(t)$ and $z_p(t)$ of a few selected particles are shown; the trajectories correspond to the particles released at different lateral ($y$-direction) locations with fixed $z = -30 \, \rm{m}$ and $x = 1490 \, \rm{m}$. Depending on their initial $y$-coordinates (marked in the middle panel of Fig.~\ref{fig:prt_trajectories}), the particles can be distinguished as the heavy-edge particles (P1 and P2), the central-region particles (P3-P5), and the light-edge particles (P6 and P7). Typically, the heavy-edge particles downwell and the light-edge particles upwell through filaments as they are transported to the central region of the front, where they circulate with the local particles organized into lobes. The lobes are associated with coherent eddies and move in the along-front direction.


The vertical trajectories of the particles are depicted in Fig.~\ref{fig:prt_trajectories}c. As expected, the figure shows that the heavy-edge particles (P1 and P2) downwell, while the light-edge particles (P6 and P7) upwell. The trajectory of P2 also shows upwelling after $t \approx 100 \, \rm{h}$, which is a result of the particle's motion within the lobe. The negative correlation between the lateral and vertical motions of the particles is also evident from some of the lateral and vertical trajectories (Figs.~\ref{fig:prt_trajectories}b and c). Such correlation often occurs for particles moving in the lobes or through the coherent filaments. For example, the $y$ and $z$ trajectories of the central-region particle P4 reveals that the particle moves vertically downward during the time when the lateral motion is in positive $y$ direction, whereas it moves vertically upward when the lateral motion is in negative $y$ direction. Moreover, the correlated lateral and vertical motions exhibit oscillations with a time period of about $25 \, \rm{h}$, which is twice the inertial time period ($T = 12.5 \, \rm{h}$). Similarly, the upwelling/downwelling edge particles (e.g. P1 and P7) exhibit negatively correlated lateral and vertical motions. The central-region particle P3 remains trapped inside an eddy and shows oscillations in its $y$ coordinates at near-inertial time scale while maintaining a nearly constant height in the vertical. The decoupled lateral and vertical motions are also evident for the light-edge particle P6, after it upwells to the surface ($t \gtrapprox 90 \, \rm{h}$). 
The vertical trajectory of P6 also shows a {fast time-scale} event with remarkably rapid transport in the vertical. This event starts at $t \approx 97 \, \rm{h}$ when P6 downwells by approximately $20 \, \rm{m}$ (from A to B) over a period of about an hour and then upwells back to the surface (from B to C) in the next two and half hours. The downwelling occurs when the particle gets attracted to a heavy-edge filament, which mainly transports cold fluid downwards. However, P6 eventually upwells when it finds itself in a denser background.  

The trajectories of the edge particles reveal that their vertical transport (under the influence of the front)  commences at different times. In general, the motion of an edge particle farther away from the central region of the front is delayed compared to the one that is closer. In Fig.~\ref{fig:prt_trajectories}c, the approximate time when the heavy- and the light-edge particles start moving vertically are marked with solid squares in their trajectories. The starting time is determined when the magnitude of vertical displacement exceeds  $2 \, \rm{m}$ for a heavy-edge particle and $4 \, \rm{m}$ for a light-edge particle; a higher threshold for the vertical displacement is used for the light-edge particles as their vertical trajectories have relatively large amplitude oscillations superposed to their initial positions. Indeed, P1 starts moving vertically after P2 and P6 after P5, since P2 and P5 are closer to the central region of the front than P1 and P6. This suggests that the vortex filaments primarily transport the edge particles adjacent to the slumping front, and those outside are transported after the width of the front increases slowly in the lateral.


The along-front particle trajectories are shown in Fig.~\ref{fig:prt_trajectories}a. The figure shows that the displacements are generally in the negative-$x$ direction, same as the mean along-front velocity at the front. {It is worth noting that the $x$ trajectories of particles P5 and P6 cross the boundary of the computational domain at $x = 0$. The trajectories are continued into the negative $x$ region using the streamwise periodicity of the domain.} It can also be noticed that the upwelling particles (P6 and P7) have larger negative $x$ displacements compared to the downwelling particles (P1 and P2), as the upwelling particles tend to spend more time near the surface where the along-front velocity is larger. Overall, the displacement of the particles in the negative $x$ direction increases with time, indicating negative along-front velocities; however, a few particles can acquire positive along-front velocities, especially when they are near the bottom of the front (e.g. P2 during $t \approx 96 - 105 \, \rm{h}$ and P4 during $t \approx 64 - 72 \, \rm{h}$)

Oscillations with small amplitudes can be observed in the vertical trajectories (e.g. P4, P5, and P6), indicating the influence of the finescales. Further, the effect of the finescales on the along-front and cross-front trajectories is weak, as the trajectories appear smooth (i.e. the deviations are small) and are dominated by the submesoscale velocity components. Next, we quantify the correlation between the lateral and vertical motions of the particles.

The particle trajectories discussed in this section have demonstrated that particles circulating in the lobes or moving through vortex filaments exhibit a trend towards negative correlation of lateral- and vertical-velocity components, a consequence of the secondary circulation induced by the coherent structures. To {\em statistically} quantify this relationship between vertical and lateral motions in the central region of the front, a correlator variable $r_{yz}$ is introduced for each particle trajectory and probability density functions (PDFs) are computed over the ensemble of particles released at specific depths. For each particle trajectory $r_{yz}$ is defined as
\begin{align} \label{eq:correlator}
 r_{yz} = \frac{\sum_{n=1}^{N} \Delta y_p^n \Delta z_p^n}{\sqrt{\sum_{n=1}^{N} (\Delta y^n_p)^2} \sqrt{\sum_{n=1}^{N} (\Delta z^n_p)^2}}.
\end{align}
Here, $\Delta y_p^n = y_p(t_{n})-y_p(t_{n-1})$ and $\Delta z_p^n = z_p(t_n)-z_p(t_{n-1})$ for a particle, and the time superscript $n$ varies to cover the entire simulation time from $t_0$ to $t_N$. The above definition of the correlator $r_{yz}$ can also be interpreted in terms of the particle velocity weighted with the advection time step, i.e., $\Delta y_p^n \approx v_p^n \Delta t_n$ and $\Delta z_p^n \approx w_p^n \Delta t_n$, with $\Delta t_n = t_n - t_{n-1}$. The weighting with $\Delta t_n$ used with the velocity components $v_p^n$ and $w_p^n$ accounts for the variable time step in the simulation.

The correlator takes a value of $r_{yz} \in [-1, 1]$, a  magnitude close to unity represent perfectly correlated vertical and lateral motions and magnitudes close to zero correspond to uncorrelated motions.
The probability density function (PDF) of the correlator $r_{yz}$ is computed for groups of particles released in the central region at different depths -- $10 \, \rm{m}$,  $20 \, \rm{m}$, $30 \, \rm{m}$ and $40 \, \rm{m}$ -- and shown in Fig.~\ref{fig:pdf_r_vw} are their PDFs. The particles quickly organize into coherent lobes after being released and and continue moving in these structures thereafter. The figure clearly shows negative $r_{yz}$ for the majority of the particles. The PDF of $r_{yz}$ for the particles released at $10 \, \rm{m}$ have a broad peak. The peak sharpens and shifts towards -1 as the depth of the release increases, indicating stronger correlation between the lateral and vertical motions. For the particles released at $40 \, \rm{m}$ depth the median value of  $r_{yz}$ is about -0.6. There are also many particles with small and moderate values of $r_{yz}$, reflecting some unpredictability in the multiscale, chaotic trajectories executed by the particles as they circulate within the front. 

{The correlator $r_{yz}$ as defined in Eq.~\ref{eq:correlator} is skewed towards the largest magnitudes of $\Delta y_p^n \Delta z_p^n$. Alternatively, we can define a correlator $\tilde{r}_{yz}$ that gives equal weight to the lateral and vertical displacements at each time step, i.e.,
\begin{equation}
 \tilde{r}_{yz} = \frac{1}{N} \sum_{n=1}^{N} \frac{\Delta y_p^n \Delta z_p^n}{|\Delta y^n_p| |\Delta z^n_p|},
\end{equation}
where $|\Delta y^n_p|$ and $|\Delta z^n_p|$ are the absolute values of $\Delta y^n_p$ and $\Delta z^n_p$, respectively. The modified correlator $\tilde{r}_{yz} \in [-1, 1]$ and contains similar information as $r_{yz}$. The PDFs of $\tilde{r}_{yz}$ for the particles released in the central region at different depths are qualitatively similar to those shown in Fig.~\ref{fig:pdf_r_vw} and are not shown. This suggests that the negative correlation between the lateral and vertical motions is not episodic, dominated by large displacements over a few time steps; instead, it is a typical feature of the particle motion within the front at all times.}   

Thus, baroclinic instability at the front leads to complex Lagrangian dynamics. The paths followed by each tracer particle vary from one another locally, as well as globally over different regions of the front. Nevertheless, an overall  underlying structure  can be constructed based on the collective motion of the particles. The tracer particles in the central region organize into lobes, each associated with a coherent submesoscale eddy. The particles in a lobe are stratified with colder particles located at the underside and warmer particles at the top. As the lobes move with eddies, particles in the lobes circulate clockwise when viewed from above, and the sense of rotation is opposite to the cyclonic coherent eddies. On average, the lateral and vertical motions are negatively correlated. Typically, a particle circulating within a lobe moves downwards when the lateral velocity is positive, but moves upwards when it becomes negative. The coherent filaments from the light/heavy side of the front connect the edges with the lobes in the central region and transfer warm/cold edge particles to the central region. {The newly deposited particles subsequently undergo the typical circulation in the lobes.} 




\section{Vertical transport} \label{sec:vertical_tranport}
In the previous section, we have shown that the vertical motions of particles at the front exhibit oscillatory components at two widely separated time scales: fast oscillations due to the small scales in the vortex filaments and slow oscillations at  a near-inertial time scale due to the circulation in the lobes.Therefore, large $w_p$ magnitudes, such as those encountered in the vortex filaments, do not necessarily lead to a net vertical transport, responsible for restratifying the front; the long-time displacements must be examined. Following a single particle over a long time is insufficient since the behavior can differ considerably from one particle to another. In this section, we investigate the collective motion of particle clouds and inquire about the relevant time scale for subduction and restratification at the front. The clouds are created such that the constituent fluid particles have similar densities and, therefore, have similar buoyancy control on the dynamics.

\subsection{Transport of particle clouds} 
Here, we investigate the vertical transport of particle clouds related to the average motion of the constituent particles. In particular, the motion of the cloud center of mass (COM) and the spread of the particles about the COM are examined. The COM of the cloud is defined as mean position of the constituent particles, i.e. $\mathbf{x}_{com}^{k} = \sum_{i = 1}^{N_k} \mathbf{x}_{i}^k/N_k$, where $\mathbf{x}_i^k$ is the position of the $i^{th}$ particle in a cloud with index $k$, and $N_k$ is the total number of particles in the cloud. The spread of the particles in the cloud is characterized by the root mean square of the particle displacement about its COM, $\mathbf{x}_{rms}^{k} = \sum_{i = 1}^{N_k} \sqrt{(\mathbf{x}_{i}^k-\mathbf{x}_{com}^k)^2}/N_k$. Figure~\ref{fig:transport_setup} shows particle clouds initially at $10\,\rm{m}$ and $40\,\rm{m}$ depths, created by dividing particles in the cross-front region $-800 < y < 800 \,\rm{m}$ into 14 groups based on their densities, as particles with similar densities are likely to have similar transport behavior. The average density (temperature) decreases (increases) progressively from cloud C1 to C14. The number of particles in each cloud and the particle distribution over the front are shown in Figs.~\ref{fig:transport_setup}a,b for the clouds released at $10\,\rm{m}$ depth and in Figs.~\ref{fig:transport_setup}c,d for those released at $40\,\rm{m}$ depth. Notice that there are more than 1000 particles in each group, giving reasonably converged statistics.

The plots of $z_{com}$ and $z_{rms}$ with time for the particle clouds are shown in Fig.~\ref{fig:vertical_transport}. First, we examine the clouds released at $10 \, \rm{m}$ depth. The plots of $z_{com}$  show subduction and upwelling of the clouds released in different regions. Typically, the clouds released over the heavy edge and the central region subduct, whereas those released at the light edge predominantly upwell. The trajectories of the subducting clouds show oscillations with near-inertial frequencies while descending to the lower depths. Among the clouds, two different types of behavior can be noticed. Clouds C2-C7 exhibit significant vertical displacement of their COM over 1-2 inertial time periods (inertial period is T = 12.5 h), which is followed by a slow adjustment. On the other hand, clouds C8-C10 show continuous subduction over the time considered here. Considering the clouds (C13 and C14) released at the light edge, C13 shows relatively weak subduction, while C14 shows upwelling. The vertical spread of the particles within the clouds is examined in Fig.~\ref{fig:vertical_transport}b, which shows $z_{rms}$ of the clouds as a function of time. The figure reveals that $z_{rms}$ curves grow over 1-2 inertial time periods and saturate to constant values, oscillating with near-inertial frequencies. The clouds released over the heavy edge and the central region (C2-C10) saturate to $z_{rms} \approx 15\,\rm{m}$, and those released over the lighter edge (C13 and C14) saturate to $z_{rms} = 5-7 \,\rm{m}$. These long-time values of $z_{rms}$ indicate the spread of the constituent particles about their centers of mass. They are larger for heavy-edge and central-region clouds than the light-edge clouds, suggesting more compact vertical configurations for the latter. {It is worth noting that the particles with similar densities remain confined, so the dispersion about the COM is primarily due to their spread about a sloping isopycnal.} 

Next, we examine the clouds released at $40 \, \rm{m}$ depth. The $z$ component of COM trajectories are plotted in Fig.~\ref{fig:vertical_transport}c. Overall, the clouds upwell, except for C2 released at the heavy edge. Among the upwelling clouds, three distinct types of behavior can be identified. First, the heavy-edge cloud C3 shows continuous rise of the COM superposed with small-amplitude near-inertial oscillations. Second, each central-region cloud (C4-C10) upwells to a peak height and then settles down to a near-equilibrium depth at  long time. Small-amplitude near-inertial oscillations can also be observed in the $z_{com}$ curves. Third, the light-edge particle clouds (C13 and C14) upwell over a longer time scale, greater than 30 h, and their long-time behavior is not clear in the present simulation. The longer time scale for the light-edge clouds is likely due a delay in the time at which most particles in the cloud start moving. Similar to the clouds released at $10 \, \rm{m}$ depth, the $40 \, \rm{m}$-depth clouds disperse vertically about their COM, as they upwell/downwell. The vertical spread of the constituent particles in the cloud $z_{rms}$ with time is shown in Fig.~\ref{fig:vertical_transport}d. The spread of the particles in the central-region clouds (C3-C10), including the particles immediately at the edges (C2 and C13), reach peak values within 1-2 inertial time periods and, subsequently, asymptote to a constant value of about $14 \, \rm{m}$. The behavior of edge particles (C2 and C14) is somewhat different. There is a delay in when the majority of the particles in the clouds is put into  motion by the coherent filaments. As a result, the peaks in $z_{rms}$ of the clouds are delayed, and their long time-behavior is not clear in the present simulation.                

The above analysis shows that the vertical transport of the edge clouds differs from the central-region clouds at both depths. Typically, the COM of the central-region clouds move vertically over 1-2 inertial time periods -- clouds released near the surface downwell while those near the bottom upwell -- and settle to a mean depth in the range of $20 \, \rm{m} - 30 \, \rm{m}$ at long times. As the clouds upwell/downwell the particles disperse about their COM. At long times, the spread of the particles about the COM saturates to a value of about $15 \, \rm{m}$. With regards to the edge clouds, their behavior is the same for both 10 m and 40 m depth of release, i.e., the heavy-edge clouds downwell and the light-edge clouds upwell. Moreover, the edge clouds exhibit slower time scales, which is likely due to the time delay over which majority of the particles in the cloud start moving. The near-inertial oscillations observed in both $z_{com}$ and $z_{rms}$ curves reflects the circulation of the constituent particles within the lobes.


The transport of fluid parcels at the front leads to its restratification. This effect of particle transport can be further elucidated by examining the PDFs of the vertical distribution of the particles. The PDFs of heavy-edge, light-edge, and central-region particles released at $30 \, \rm{m}$ depth are considered separately, as shown in Fig.~\ref{fig:vertical_distribution_pdf}. In this figure, the  PDFs are plotted at t = 78.5 h, 86 h, and 95 h, which correspond to the particle flight times of about $\Delta t = 20 \, \rm{h}, \, 30 \, \rm{h},$ and $50 \, \rm{h}$ after the release. They are constructed by dividing the domain into horizontal slabs of $2 \, \rm{m}$ thickness and sampling the particles in them. We find that within $\Delta t = 20 \, \rm{h}$, the vertical distribution of the central-region particles reaches a quasi-steady profile that changes slowly with time. At the heavy edge, particles downwell through filament structures to the central region. At t = 78.5 h, most of the particles remain at $z \approx -30 \, \rm{m}$, and a large peak is observed at this depth. As time progresses, the peak reduces in height and the probability corresponding to $z < -30 \,\rm{m}$ increases, indicating downwelling and subduction of the heavy-edge particles. In contrast, the light edge particles upwell. There is a large peak near $z = -30 \, \rm{m}$ depth at $t = 78.5 \, \rm{h}$ as most of the particles remain uninfluenced by the instabilities. However, the peak reduces in height at later times as the particles upwell, and the probability with $z > -30 \, \rm{m}$ grows. The whole process can be summarized as follows. The coherent structures at the front quickly, over a period of about 20 h,  distribute the central-region particles vertically into a nearly stable configuration. This is a consequence of the fast dynamics inherent in the system. Subsequently, the distribution changes slowly when edge particles are drawn into the central region through the coherent filaments. The flux of new particles into the central region causes the particles already present in the region to adjust and the front restratifies. 


\subsection{Transport of fluid and  flow properties}
As Lagrangian particles move with the fluid, they carry the properties associated with  material points, e.g., fluid properties such as temperature and flow properties such as kinetic energy (KE). These properties may change due to turbulent exchanges and dynamical interactions with the surrounding fluid. For example, KE can change because of subgrid and viscous diffusion, as well as pressure and buoyancy interactions. The overall changes in flow properties have important implications for the final state of the front and also for understanding the subduction of surface properties to the bottom of the surface layer. Here, we investigate the average subgrid viscosity experienced by the cloud particles, reflecting turbulent mixing with surrounding fluid, and also changes in temperature and KE as the clouds C1-C14 are transported by the submesoscale currents.

The exchange of flow properties between a fluid particle and its surroundings depends on the local gradient of the property, as well as the turbulence characterized here by the subgrid viscosity. Note that filaments have high levels of the turbulent finescale and subgrid viscosity. In Fig.~\ref{fig:transport_nut}, the average subgrid viscosities experienced by the particle clouds released at $10\, \rm{m}$ and $40 \, \rm{m}$ depths are plotted as a function of time. It can be observed in Fig.~\ref{fig:transport_nut}a, showing $10\, \rm{m}$-depth particle clouds, that the heavy-edge and the central-region clouds, especially when they are near the surface, experience larger subgrid viscosities than the  clouds at the light edge. The magnitudes reduce as they subduct further down below the surface. In contrast, the upwelling clouds (C13 and C14) experience relatively weaker subgrid viscosities, which remain nearly constant with time. The large magnitudes of subgrid viscosity sampled by heavy-edge and central-region clouds are associated with downwelling through vortex filaments where strong finescales are present (VPS2019). The finescale is generated through frontogenesis and is particularly energetic near the surface, where frontogenesis is intensified \citep{LapeyreKH:2006}. Thus, particles get attracted to the coherent filaments as they downwell/upwell at the front. Furthermore, the finescale and its associated subgrid viscosity is weaker in these structures at depth. The  mean value of  subgrid viscosity experienced by the clouds lies in the range of $200\nu$ to $300\nu$, when the clouds are near the surface. These values reduce as the clouds subduct, and at late times the mean subgrid viscosity experienced by the clouds becomes $\sim 80\nu$. 

The magnitudes of mean subgrid viscosity experienced by upwelling clouds (C13 and C14) are smaller initially, but asymptote to values comparable to those attained by the heavy-edge and the central region clouds at long times. The average subgrid viscosities experienced by the particle clouds released at $40 \, \rm{m}$ depth are depicted in Fig.~\ref{fig:transport_nut}b. The figure shows higher magnitudes of mean subgrid viscosities for the central-region clouds initially as they get attracted to the upwelling filaments, but the magnitudes are about two-third of the corresponding values of the $10 \, \rm{m}$-depth clouds. This suggests weaker frontogenesis and finescales at depth. Moreover, as the particles upwell the mean subgrid viscosity experienced by the clouds become smaller. In some upwelling clouds (e.g., C9 and C10), elevated mean subgrid viscosities can be observed at intermediate times($t \approx 25 - 35 \, \rm{h}$) which is due to trapping of the particles by the downwelling filaments near the surface. Initially, the edge particles (C2, C13 and C14) are away from the filaments and show weaker mean subgrid viscosities, but the magnitudes increase when they upwell/downwell through vortex filaments. 

Because of turbulent diffusion, contact with the surrounding fluid changes the mean temperatures of particle clouds. In Fig.~\ref{fig:transport_rho}, the deviation of the mean temperature from the initial mean value normalized by the across-front temperature difference $\Delta_F T$ is depicted for each of the considered particle clouds. The mean of temperature change for the clouds released at $10 \, \rm{m}$ depth are plotted with time in Fig.~\ref{fig:transport_rho}a. From the figure, it can be observed that clouds released at the heavy edge and those in the neighborhood become warmer with time -- the mean temperature of C2 and C3 rises continuously, while C4 and C5 become warmer at late times (after about $\Delta t = 30 \, \rm{h}$). The mean temperature of the remaining clouds decreases with time, and they become relatively heavier as they downwell/upwell. Similar trends are observed for the particle clouds released at $40 \, \rm{m}$ depth. {One striking difference can be noticed with the heavy-edge cloud C2 at 40 m depth, which becomes colder as opposed to becoming warmer. The reason is that the particles in cloud C2 come in contact with colder thermocline water that is pulled into the cyclonic eddy due to the eddy suction.} It can also be noted that the changes in the mean temperatures of the $40 \, \rm{m}$-depth clouds are smaller compared to those released at $10 \, \rm{m}$ depth. This correlates with the average subgrid viscosities experienced by these clouds, with magnitudes generally being larger for the $10 \, \rm{m}$-depth clouds. Overall, during the time (approximately 45 hrs)  of the particle advection, the mean temperatures of the clouds change by $4-6\%$ with respect to the imposed lateral temperature difference across at the front.

We note that the net change in mean cloud temperature during the  the entire advection time of $\tau_e = 56 \, \rm{h}$ is primarily due to subgrid diffusive processes reflected by $\nu^{sgs}$; the contribution of molecular diffusion acting on the horizontal and vertical gradients of temperature is relatively much smaller. The change in temperature of a particle resulting from the molecular diffusion of the horizontal temperature gradient can be estimated as $\kappa (\partial^2 T_p/\partial y^2) \tau_e \approx 0.14\nu (\Delta_F T/W_F^2) \tau_e$, where $\Delta_F T = 0.09 \, \rm{K}$ is the temperature difference across the front and $W_F \sim 100 \, \rm{m}$ is the width of the vortex filaments. Thus the change in temperature of the particle is approximately $0.3\Delta_F T \times 10^{-5} K$, about three orders of magnitude smaller than the values observed here. Similarly, the change in particle temperature because of the vertical diffusion can calculated as $0.14\nu (\partial^2 T_p/\partial z^2) \tau_e \approx 0.14\nu (\nabla_F T/H^2)\tau_e$, and since $H = 50 \, \rm{m}$, the temperature change is about four times larger than the horizontal counterpart, but the net contribution is still not significant. 

The ensemble-averaged KE of the $10\,\rm{m}$-depth and $40\,\rm{m}$-depth clouds is plotted as a function of  time in Figs.~\ref{fig:transport_ke}a,b. As the clouds upwell or downwell, the mean KE changes. Typically, the downwelling clouds lose KE, whereas upwelling clouds gain KE (Fig.~\ref{fig:transport_ke}). This behavior suggests a prevalence of an overall balance in the dynamics at the front that results in decreasing KE with depth. The KE plots also reveal near-inertial oscillations that are quite significant in the central-region clouds released at $10 \, \rm{m}$ depth (e.g., C7 in Fig.~\ref{fig:transport_ke}a). The near-inertial oscillations in KE can be attributed to the circulation of the particles in the lobes. 

%
%
\section{Dispersion} \label{sec:dispersion}
In this section, single- and two-particle dispersion statistics, as well as multiparticle statistics using a group of four particles (tetrads) are studied. All the results included in this section consider only those particles released in the central region. 

\subsection{Single-particle dispersion}
Single-particle dispersion, also known as absolute dispersion, is calculated as the mean square displacement over an ensemble of particles. Thus, absolute dispersion is given by
\begin{equation} \label{eq:absolute_dispersion}
 A^2(t) = \langle (\mathbf{x}(t)-\mathbf{x}(0))^2 \rangle,
\end{equation}
where $\mathbf{x}(t)-\mathbf{x}(0)$ is the displacement of a tracer particle and $\langle \cdot \rangle$ represents the mean taken over the particle ensemble. The expression of absolute dispersion in Eq.~\ref{eq:absolute_dispersion} can be expressed as $A^2(t) = A_x^2(t) + A_y^2(t) + A_z^2(t)$, with $A_x^2(t)$, $A_y^2(t)$ and $A_z^2(t)$ representing contributions from displacements along $x$, $y$ and $z$ directions, respectively.  

Unlike homogeneous isotropic turbulence, particle dispersion is anisotropic in this problem since the particles move in a stratified environment under the action of coherent structures and a mean downfront jet with vertical and lateral shear. In Fig.~\ref{fig:absolute_dispersion}, the absolute dispersion components $A_x^2(t)$, $A_y^2(t)$ and $A_z^2(t)$ of the particles released at $10 \, \rm{m}$ and $30 \,\rm{m}$ depth are plotted. Initially, particles disperse ballistically and each of the three components grow as $t^2$. The vertical dispersion $A_z^2(t)$ starts to deviate from $t^2$ behavior at $t \approx 0.5 \, \rm{h}$, when the root-mean-square (rms) displacement in the vertical is about $2 \, \rm{m}$, and at late times, it saturates to an rms value of $O(10) \, \rm{m}$. The dispersion components in $x$ and $y$ directions grow as $t^2$ over longer time durations. The long-time behavior, on the other hand, is super-diffusive in the $x$ direction with $A_x^2(t)$ growing as $t^{1.8}$ and diffusive in the $y$ direction with $A_y^2(t)$ growing as $t$. 

At late times, diffusive behavior is commonly anticipated since the motions of the particles moving under the influence of different eddies become uncorrelated. The observed super-diffusive behavior in $A_x^2(t)$ can be attributed to the horizontal and vertical shear of the mean along-front velocity. By using simple stochastic models, it can be demonstrated that in a sheared velocity field absolute dispersion can grow as $t^{\alpha}$, where $1 \le \alpha \le 3$, and the value of the exponent $\alpha$ depends on the shear profile \citep{LaCasce:2008}. For example, if the velocity is in the $x$ direction with a constant shear along the $y$ direction, then the random walk by advecting particles in the direction of the shear produces absolute dispersion whose $x$ component grows as $t^3$. Supper-diffusive behavior is observed in other flow configurations as well. In stratified turbulence, this behavior arises due to  vertical shear of the horizontal velocities in the layers between coherent pancake eddies \citep{vanAartrijkCW:2008}.

It is worth noting that the behavior of $A_x^2(t)$, $A_y^2(t)$ and $A_z^2(t)$ is qualitatively similar for the groups of particles released at $10\,\rm{m}$ and $30 \, \rm{m}$ depth. However, some differences can be noticed in the growth of $A_x^2(t)$, which is initially considerably faster for the particles released at 10 m depth compared to those released at 30 m depth. At late times, the dispersion curves tend to converge. This behavior can be understood considering the vertical shear of the along-front velocity: the velocity component is strongest near the surface, but decreases with depth and becomes zero near the bottom of the mixed layer. At late times, the particles released at both depths become vertically dispersed, and those advecting near the surface dominate the super-diffusive growth of $A_x^2(t)$, leading to similar dispersive behavior.


\subsection{Particle-pair dispersion}
Pair dispersion, also known as relative dispersion, is calculated as the mean-square pair separation. Thus, relative dispersion is expressed as
\begin{equation} \label{eq:relative_dispersion}
 R^2(t) = \langle (\mathbf{x}^{(1)}(t)-\mathbf{x}^{(2)}(t))^2 \rangle,
\end{equation}
where $\mathbf{x}^{(1)}(t)$ and $\mathbf{x}^{(2)}(t)$ are the positions the particles in a pair, and $\langle \cdot \rangle$ represents the mean taken over all the selected pairs. From Eq.~\ref{eq:relative_dispersion}, $R^2(t) = R_x^2(t) + R_y^2(t) + R_z^2(t)$, with $R_x^2(t)$, $R_y^2(t)$ and $R_z^2(t)$ representing contributions from the relative displacements of the particle pairs along $x$, $y$ and $z$ directions.

Relative dispersion also signifies the spread of a cloud of particles about the center of mass (COM). The short- and long-time behaviors of pair-dispersion are easily understood. For short times, the difference between the velocities of the particles in a pair is nearly constant since the particles are nearby, and the mean square of pair separation grows ballistically as $t^2$. On the other hand, at long times, the pairs become widely separated, so that the motions of the particles in a pair are influenced by different eddies and become uncorrelated. These pairs with uncorrelated motions lead to a long-time pair-dispersion behavior that is similar to single-particle dispersion. 

It is the intermediate time- and length-scale behavior of the pair dispersion that is of interest since they reveal the internal dynamics of the flow. When the pair separation is in the inertial range of a forward energy cascade, the application of Kolmogorov similarity hypothesis suggests $R^2(t) \sim \varepsilon t^3$, where $\varepsilon$ is the rate of KE dissipation. However, the similarity hypothesis can be applied only at length scales which are much larger than the viscous dissipation scale and are also sufficiently small to remain unaffected by external influences and the boundary. Oceanic flows are constrained by rotation and stratification, which leads to quasi-2D flows at sufficiently large scales, with horizontal velocity magnitudes much larger than the vertical. Turbulence generated in such flows behave differently and exhibits two inertial ranges \citep{Kraichnan:1967, Charney:1971}: a forward cascade of enstrophy to smaller scales and a backward cascade of energy to larger scales. The two cascades start in the neighborhood of the scale where external forcing is applied. Applying the similarity analysis to the regime of forward enstrophy cascade, the pair dispersion can be expressed as $\exp (c_3 \eta^{1/3} t)$, where $\eta$ is the rate of enstrophy cascade \citep{Lin:1972}. For the regime of the inverse energy cascade we again get the same expression as the forward energy cascade, but $\varepsilon$ here represents the rate of energy transfer to the larger scale. {There is evidence of forward enstrophy cascade in the ocean at length scales below the deformation radius, e.g. the central part of the North Atlantic \citep{OllitraultGD:2005} and the Gulf of Mexico \citep{BalwadaLS:2016}. In 2D homogeneous and isotropic turbulence, the exponential growth of the relative dispersion can be associated with  non-local dynamics whose energy spectra varies as $E(k) \sim k^\beta$ with $\beta \ge 3$ \citep{Bennett:1984}.  

The time evolution of the relative dispersion of the particle pairs released in the central region of the front, $-500 \, \rm{m} < y < 500 \, \rm{m}$, at $10\, \rm{m}$ and $30 \, \rm{m}$ depth  is shown in Fig.~\ref{fig:relative_dispersion}a and $10 \, \rm{m}$ depth and the surface in Fig.~\ref{fig:relative_dispersion}b. Following the approach used for examining absolute dispersion, the three contributions to relative dispersion $R_x^2(t)$, $R_y^2(t)$ and $R_z^2(t)$ are investigated separately. Different pairs are considered for calculating different components: the nearest neighbors separated in $y$ direction are considered for $R_x^2(t)$, those separated in $x$ direction for $R_y^2(t)$, and the nearest neighbors in both $x$ and $y$ directions are considered for $R_z^2(t)$.

First, we focus on the particles released at $10 \, \rm{m}$ depth (solid lines in Fig.~\ref{fig:relative_dispersion}a). All the three components of relative dispersion grow as $t^2$ in the beginning. The vertical component of the relative dispersion starts to deviate from $t^2$ at $t \approx 0.4 \, \rm{h}$, when the rms of the pair separation in the vertical is $\sim2 \, \rm{m}$, much smaller than the depth of the mixed layer. Subsequently, the vertical component of the pair dispersion grows slowly and finally saturates at O(10) m, which is similar to the square root of the absolute dispersion in the vertical $|A_z(t)|$ at late times. The horizontal components of the relative dispersion $R_x^2(t)$ and $R_y^2(t)$ show ballistic growth over a longer time duration, up to $\Delta t \approx 2\,\rm{h}$. The corresponding rms pair separation is O(10) m, with the magnitude being slightly larger for the $x$-component, $R_x^2(t)$. As previously explained, the late-time behaviors of the horizontal components $R_x^2(t)$ and $R_y^2(t)$ are same as those obtained for the corresponding single-particle dispersion: $R_x^2(t)$ shows super-diffusive behavior with the mean-square pair separation growing as $t^{1.8}$, while $R_y^2(t)$ shows diffusive behavior with the mean-square pair separation growing as $t$. During the intermediate times, the horizontal components $R_x^2(t)$ and $R_y^2(t)$ exhibit $t^3$ growth. This may indicate a possible inertial range with forward energy cascade at the intermediate scales. We further note that the rms of the relative displacements of the particle pairs during the intermediate times is $O(100)\, \rm{m}$, which is comparable to the width of the vortex filaments. 

The relative dispersion of the particle pairs released at $30 \, \rm{m}$ depth behaves qualitatively similar to the pairs released at $10 \, \rm{m}$ depth, and the curves of the relative-dispersion components follow closely for the two groups of particles. Nevertheless, the initial growth of the dispersion components is somewhat smaller for the particles released at $30 \, \rm{m}$ depth compared to those released at $10 \, \rm{m}$ depth; however, the differences become smaller at late times. 

In contrast to the behavior of horizontal components $A_x^2(t)$ and $A_y^2(t)$ during the initial and intermediate times, the relative dispersion components $R_x^2(t)$ and $R_y^2(t)$ are comparable to each other for both $10\,\rm{m}$- and $30\,\rm{m}$-depth particles. Hence, although the frontal jet is aligned with the $x$-direction, the relative motions of the particles in the horizontal plane are isotropic. However, relative dispersions in the $x$ and $y$ directions diverge at late times. It can be attributed to the fact that the domain is infinitely long in the along-front direction, but it is confined in across-front direction. As a result, the pair separation in the along-front can grow to become much larger than that in the across-front.

The growth of $R_x^2(t)$ and $R_y^2(t)$ as $t^3$ during intermediate times does not necessarily imply the existence of an inertial range with forward energy cascade. There are dynamics fundamentally different than inertial-range 3D turbulence, which can produce this behavior, e.g., shear dispersion. To further investigate the intermediate-scale dynamics, we examine the relative dispersion of the particle pairs released at the surface. In Fig.~\ref{fig:relative_dispersion}(b), $R_x^2(t)$ and $R_y^2(t)$ are compared between  surface particles and those released at $10 \, \rm{m}$ depth. At the surface, vertical velocity is imposed to be zero and the flow is essentially 2D, with the particles constrained to move in the horizontal plane. Interestingly, the surface pairs also exhibit $t^3$ growth of $R_x^2(t)$ and $R_y^2(t)$ during the intermediate times. However, visualization of the motions of the surface particles reveals dispersion by the horizontal shear and straining in the vortex filaments, where the particles get attracted after their release. The influence of  the turbulent finescale on relative dispersion in the horizontal is weak and is dominated by the  energetic submesoscale flow (including the mean). Indeed, examining the $x$ and $y$ components of the particle trajectories reveals the horizontal motions controlled mainly by the large-scale component, so much so that the influence of the finescale is negligible on horizontal trajectories, as was illustrated for particles released at $30 \, \rm{m}$ in Fig.~\ref{fig:prt_trajectories}. It is worth noting that the surface pairs transition to the long-time behavior earlier than those released at $10 \, \rm{m}$ depth. During this phase, both $R_x^2(t)$ and $R_y^2(t)$ grow as $t^{1.8}$. However, since the front is of finite width, $R_y^2(t) \sim t^{1.8}$ growth cannot be maintained over a long time duration. Ultimately, $R_y^2(t)$ is likely to saturate to a growth in time reflective of the widening of the front.

{We also note that the submesoscale turbulence simulated here does not show any evidence of  exponential growth for the relative dispersion components in $x$ and $y$ directions, below the deformation radius, which is comparable to the diameter of the submesoscale eddies. The observed relative dispersion of $R_x^2(t)$ and $R_y^2(t)$ is consistent with the fact that the energy spectra of the velocity $E(k) \sim k^{-\beta}$, where the exponent $\beta$ lies in the range 2-3. For nonlocal dispersion with exponential growth of pair separation, the 2D flows are required to have $\beta \ge 3$.}

\subsection{Multiparticle dispersion}
In turbulent flows, a cluster of fluid particles is strained by correlated large-scale motions. Additionally, the constituent particles disperse randomly due to independent and incoherent finescale turbulence. The large-scale motions can lead to the deformation of the cluster into flow-specific geometries, whereas finescale fluctuations lead to an increase in the average volume while maintaining the overall shape. \citet{PumirSC:2000} introduced a statistical measure using three or more material points to probe the geometry of Lagrangian dispersion. Here, we investigate the shape changes by tracking groups of four particles. Following \citet{PumirSC:2000},  the geometry of the tetrad is defined by the following three vectors:
\begin{align}
 \mathbf{r}_1 &= \frac{1}{\sqrt{2}} (\mathbf{x}_p^{(1)} - \mathbf{x}_p^{(2)}), \\
 \mathbf{r}_2 &= \frac{1}{\sqrt{6}} (2\mathbf{x}_p^{(3)}- \mathbf{x}_p^{(1)} - \mathbf{x}_p^{(2)}), \\
 \mathbf{r}_3 &= \frac{1}{\sqrt{12}} (3\mathbf{x}_p^{(4)}- \mathbf{x}_p^{(1)} - \mathbf{x}_p^{(2)} - \mathbf{x}_p^{(3)}),
\end{align}
where $\mathbf{x}_p^{(i)}$ with $i = 1, \,2, \,3, \,4$, are the position vectors of the four particles at the vertices of a tetrahedron. The radius of gyration of the cluster is $R^2 = \sum_{i=1}^3 \mathbf{r}_i^2$ and measures the spatial extent of the tetrad. The vectors involving position differences are combined into a second order tensor 
\begin{align}
 \mathbf{g} = \mathbf{r}\mathbf{r}^{t},
\end{align}
where $\mathbf{r} = [\mathbf{r}_1, \, \mathbf{r}_2, \, \mathbf{r}_3]$ is a second order tensor with $\mathbf{r}_1$, $\mathbf{r}_2$, and  $\mathbf{r}_3$ as its column vectors. The eigenvalues of $\mathbf{g}$ ($g_1 > g_2 > g_3$) provide a convenient characterization of the shape of the particle cluster. For example, $g_1 = g_2 = g_3$ corresponds to an isotropic object, $g_1 \approx g_2 \gg g_3$ corresponds to a pancake-like object which has much smaller vertical scale compared to the horizontal, and $g_1 \gg g_2, \, g_3$ corresponds to a needle-like object. The eigenvalues are often normalized by the radius of gyration $R^2 = \text{Trace}({\mathbf{g}})$, i.e. $I_i = g_i/R^2$, in order to facilitate comparison of shapes at different times. The multiparticle statistical measure described above has been used in the studies of homogeneous isotropic turbulence and stably stratified homogeneous turbulence to understand the Lagrangian shape dynamics. The present study is an application of this multiparticle measure to a flow with submesoscale currents. 

{For the multiparticle study, two  particles were added around each  node of the particle-lattices at $10\,\rm{m}$ and $30\,\rm{m}$ depth. A tetrad was formed with the nodal particle, the two additional particles, and the particle above it in the original lattice.
The construction of the tetrads is illustrated in Fig.~\ref{fig:tetrad_construction} by visualizing the tetrads in a small patch. In this figure, the blue particles represent the node particles, while the red and the green particles are the added particles, placed 2 m apart in the $x$ and $y$ directions, respectively, from the blue nodal particles. The black particles are the node particles one level above the base level ($10 \, \rm{m}$ and $30 \, \rm{m}$ depth), i.e., $2 \, \rm{m}$ above the base level.} 

The results presented in this section include only those tetrads released in the central cross-front region, i.e., $-500\,\rm{m}<y<500\,\rm{m}$. The normalized eigenvalues $I_1$, $I_2$ and $I_3$, averaged over the tetrads, are plotted as a function of time in Fig.~\ref{fig:multiparticle_dispersion}a. The figure shows rapid deformation of the tetrahedra into flattened, needle-like objects {\em within an hour after  release}  as $\langle I_2 \rangle$, $\langle I_3 \rangle \approx 0$ and $\langle I_1 \rangle \approx 1$. Subsequently, $\langle I_2 \rangle$ plateaus during $\Delta t = 1 - 10 \,\rm{h}$ and increases slightly at late times. $\langle I_3 \rangle$, on the other hand, continues to drop. During the time interval when $\langle I_2 \rangle$ plateaus, the horizontal components of the pair separation $R^2_x(t)$ and $R^2_y(t)$ are observed to transition from the short-time $t^2$ dispersion regime to the long-time super-diffusive dispersion regime, as particles move through the filament structures. At long times,  $\langle I_2 \rangle$ tends to approach a constant value, but its magnitude remains considerably smaller than that of $\langle I_1 \rangle$.

The average values of $I_1$, $I_2$ and $I_3$ show the predominance of flattened, needle-like objects by $\Delta t \approx 1\,\rm{h}$ after the release of the tetrads. It is possible that other shapes are also present. In order to evaluate the distribution of shapes, the  PDFs of $I_1$ and $I_2$ are plotted in Fig.~\ref{fig:multiparticle_dispersion}b at different times after the release of the particle clusters. It can be seen from the figure that within $\Delta t = 20 \, \rm{min}$, the peak in the PDF of $I_1$ shifts to values greater than 0.5 and that of $I_2$ to values smaller than 0.5. However, there are a few tetrads which can be considered pancake-like. After $\Delta t = 40 \, \rm{min}$, distinct peaks appear for $I_1$ and $I_2$ close to 0.9 and 0.1, respectively. As time progresses, the peak at 0.9 moves towards 1 and that at 0.1 moves towards 0 as is evident upon comparison of the PDF at  $\Delta t = 1.2\,\rm{h}$ with the PDF at $\Delta t = 40 \, \rm{min}$. Thus, most of the clusters deform into primarily flat, needle-like objects by $\Delta t = 1.2\,\rm{h}$. Even at late times, the PDFs of $I_1$ and $I_2$ do not change significantly and the needle-like shapes remain dominant. This indicates dominance of the larger-scale submesoscale currents over the turbulent finescale in the present problem. Visualization of particles reveals that particles are attracted to the coherent filaments after they are released. The  high strain rates found  in the filaments and at the outer edges of the submesoscale eddies act on the clusters to deform them into needle-like objects.

The overall long-time shape distribution observed in the present problem with submesoscale currents and the finescale organized in vortex filaments can be compared with that observed in homogeneous, isotropic turbulence and in stratified turbulence. In homogeneous and isotropic turbulence, the motions of the particles become uncorrelated at long times and the ensemble averages of $I_1$, $I_2$ and $I_3$ converge to constant values consistent with the Gaussian distribution of the particles: $\langle I_1 \rangle_G \approx 0.748$, $\langle I_2 \rangle_G \approx 0.222$ and $\langle I_3 \rangle_G  \approx 0.03$. The small value of $\langle I_3 \rangle_G$ compared to $\langle I_1 \rangle_G$ and $\langle I_2 \rangle_G$ suggests that the shapes are dominated by flat objects \citep{PumirSC:2000, BiferaleBCDLT:2005}. In stratified turbulence, where vertical motions are suppressed by stratification, the final shape depends on the strength of vertical stratification, measured by the buoyancy frequency  $N$.  \cite{vanAartrijkCW:2008} found that $\langle I_1 \rangle$ becomes larger and $\langle I_2 \rangle$ smaller as stratification grows stronger. In their study with strong stratification (N100 with $N = 0.98 \, \rm{s^{-1}}$) the shapes overall were  needle-like, but the shape distribution examined by plotting the PDFs of $I_1$ and $I_2$ revealed the presence of a significant number of  flat objects. In the present work, the long-time average values are $\langle I_1 \rangle \approx 0.93$, $\langle I_2 \rangle \approx 0.07$, and $\langle I_3 \rangle \approx 0$, suggesting predominantly flat, needle-like shapes. Further, the shape distribution examined using PDFs of $I_1$ and $I_2$ shows that most of the tetrahedra are deformed into such objects.    

\section{Discussion and conclusions} \label{sec:discusion_conclusion}
We investigate dispersion and transport by submesoscale turbulent currents generated by the evolution of baroclinic instability at an upper-ocean front. The study employs a LES model and is performed in the Lagrangian framework by releasing a large number of tracer particles that move with the local fluid velocity. The presence of coherent structures such as vortex filaments and eddies is a typical feature of submesoscale dynamics. From the Lagrangian analysis, we find that these structures provide the primary pathways of three-dimensional transport by submesoscale currents and provide a quantitative assessment of the transport.  


The paths followed by individual particles are found to be complex and can differ considerably as time progresses, even for a pair released close to each other. Nevertheless, the motions are strongly influenced by the coherent structures, namely the vortex filaments and eddies. {Particles inside the filaments experience rapid motions with displacements of $O(10) \, \rm{m}$ over an hour, as well as slower motions at a near-inertial time scale while moving under the influence of the coherent structures.} It is possible to identify some common features that dictate the overall transport. 
The central-region particles cluster into inclined {\em lobes}, each associated with a coherent eddy. The lateral and vertical velocity of these particles reveals a clockwise circulation when viewed from above, which is opposite to the circulation {induced by} the coherent cyclonic eddies. The vortex filaments connect the heavy and light edges of the front with the central region and play a critical role in vertical and lateral transport and {the restratification of the front}. The process can be described as follows. The coherent filaments draw the edge particles into the central region and transfer them to the lobes. The lobes are stratified, and the heavy-edge particles downwell to the undersides of the lobes, whereas the light-edge particle upwell to the top. The flux of new edge particles into the central region from the edges causes the central-region particles to adjust, which leads the front to restratify. 

We find that the lateral ($v_p$) and vertical ($w_p$) velocity of the particles moving through the filaments and circulating in the lobes have a near-inertial time scale of $O(2\pi/f)$ and have a correlation which is consistent {with the lateral stratification}. For the present case where {the cross-front density gradient} is negative, the correlation is also negative. The correlation between $v_p$ and $w_p$ is quantified  by defining a correlator {for each particle using its lateral and vertical displacements over small time intervals and computing the accumulated value over the entire flight time of the particle.} The median value of the correlator determined for the central region particles is $\sim - 0.5$; the value depends on the depth where the particles are released with the correlation being somewhat larger for particles released near the bottom. It is important to note that the large magnitudes of vertical velocity in the filaments {or lobes do not} independently lead to a net restratification of the front. The restratification process is also dependent on the transport of particles from the edges to the central region with an appropriate correlation between $v_p$ and $w_p$.

Further analysis by following the centers of mass of the particle clouds $z_{com}$  released at $10 \, \rm{m}$ and $40 \, \rm{m}$ depth shows that subduction/upwelling through the vortex filaments occurs over 1-2 inertial time periods, and a slow adjustment follows {after the particles are accommodated in the lobes and begin circulating.} The near-inertial time scale is consistent with the time scale of the growth of baroclinic instability, which drives the restratification of the front. During the subduction/upwelling through the filaments, the particles disperse, mostly along the sloping isopycnals, and the the root-mean-square vertical displacement ($z_{rms}$)  of the constituent particles with respect to the COM  saturates to $\sim 15 \, \rm{m}$. Near-inertial oscillations in $z_{com}$ and $z_{rms}$ of the particle clouds are observed to result from the  circulation  of the particles in the lobes. 

Fluid and flow properties associated with  material points are also transported by the submesoscale currents. The mean subgrid viscosity of the particle clouds released at  $10 \, \rm{m}$ and $40 \, \rm{m}$ depth reveals large magnitudes initially, reflecting the motions of the particles through the vortex filaments. Moreover, the clouds released near the surface experience about 2-3 times larger values of subgrid viscosity compared to those released  near the bottom, as the finescale activity near the surface is stronger due to surface-intensified frontogenesis. The mean values typically decrease for both groups of particle clouds as time progresses. The possible exceptions are the {upwelling} particle clouds, which show elevated mean subgrid viscosity at late times when they reach the near-surface region. The average subgrid viscosity at late time is about $50-100$ times larger than the molecular viscosity for both $10\,\rm{m}$-depth and $40\,\rm{m}$-depth clouds. Because of the turbulent exchange with the surroundings, the fluid properties associated with the particles change. We find that the change in the average temperature of the clouds over a flight time of about 45 h is about $4-6\%$ of the cross-front temperature difference. Typically, the downwelling particles on average tend to become warmer while the upwelling particles tend to become colder. The mean KE of the clouds change with $z_{com}$, which reflects a more energetic flow near the surface than at depth.

{The process of restratification in the present model front under BI is considerably different than the restratification process depicted in \cite{Spall:1995}, which assumes sliding of a fluid parcel from the heavy side of the front across to the light side. In contrast, the process described here is three-dimensional and involves continuous stirring of the central-region fluid by submesoscale coherent eddies and the injection of edge particles into the central region by the coherent vortex filaments. Thus, after being subducted, a heavy-edge fluid parcel continues to move under the influence of the eddies at the front.

We also find that vertical distribution of the particles released at a depth remains confined within $\sim 50 \, \rm{m}$ depth from the surface over time (see Fig.~\ref{fig:vertical_distribution_pdf}), which is also the initial depth of the front. As shown in VPS19, the vertical velocity in the thermocline is non-zero, but the particles do not subduct below the surface layer. This behavior is consistent with the fact that coherent structures control the vertical transport of the particles at the front. Since these structures are contained within the front, so are the particles. }

The near-inertial oscillations observed here can be contrasted with the inertial oscillations associated with the geostrophic adjustment of a front with an initially unbalanced horizontal density gradient, which was analyzed by \citet{TandonG:1994}. In the present simulation, the near-inertial oscillations result primarily from the anticyclonic circulation of the particles within the lobes. Furthermore, the dynamics here are driven by BI. In contrast, the model investigated by \citet{TandonG:1994} exhibits inertial oscillations when the unbalanced front tries to slump by releasing potential energy, but the Coriolis force acts on the developed velocity field and provides a restoring tendency towards the original configuration. \citet{TandonG:1994} find that the oscillatory adjustment  continues at inertial time scale and, due to the lack of dissipation in their simplified model,  the system does not return to a stable stationary state.

The dispersion characteristics of the submesoscale turbulent flow are also studied here. In both single- and pair-particle dispersion, the vertical component is restricted by the mixed layer depth and its value saturates to $O(10)\,\rm{m}$ at long times. The along-front component of single-particle dispersion shows super-diffusive behavior at late times, and the mean-square displacement increases as $t^{1.8}$; this behavior can be related to the mean jet in the negative $x$ direction. In the ocean,  long-time super-diffusive behavior has been observed in  coastal regions with mean currents. The particle-pair dispersion in $x$ and $y$ directions show $t^3$ behavior during the intermediate times, and the root-mean-square displacement is $O(100) \, \rm{m}$, {which is comparable to the lateral width of the vortex filaments}. This may indicate a Kolmogorovian inertial range,
 but the role of {horizontal shear} on relative dispersion cannot be ruled out. The long-time behavior of particle pairs is consistent with  single-particle dispersion. The multiparticle analysis reveals strong filamentogenesis in the vortex filaments, as the tetrads moving through these structures deform into thin, needle-like shapes. Probability density functions of {shape metrics} $I_1$ and $I_2$ indicate that there is a strong propensity to form needle-shaped structures, more so than in homogeneous turbulence that is either isotropic or stratified. The filamentogenesis is associated with the strong strain field within the coherent filaments.

\bigskip\noindent{\bf Acknowledgments}

We are pleased to acknowledge the support of ONR grant N00014-18-1-2137. We thank Hieu Pham for helpful discussion and his comments on a draft version of this manuscript. 




\bibliographystyle{elsarticle-harv}
\bibliography{references}


%
%
%

\pagebreak
\begin{figure}[!ht]
 \centering
 \includegraphics[width=\linewidth]{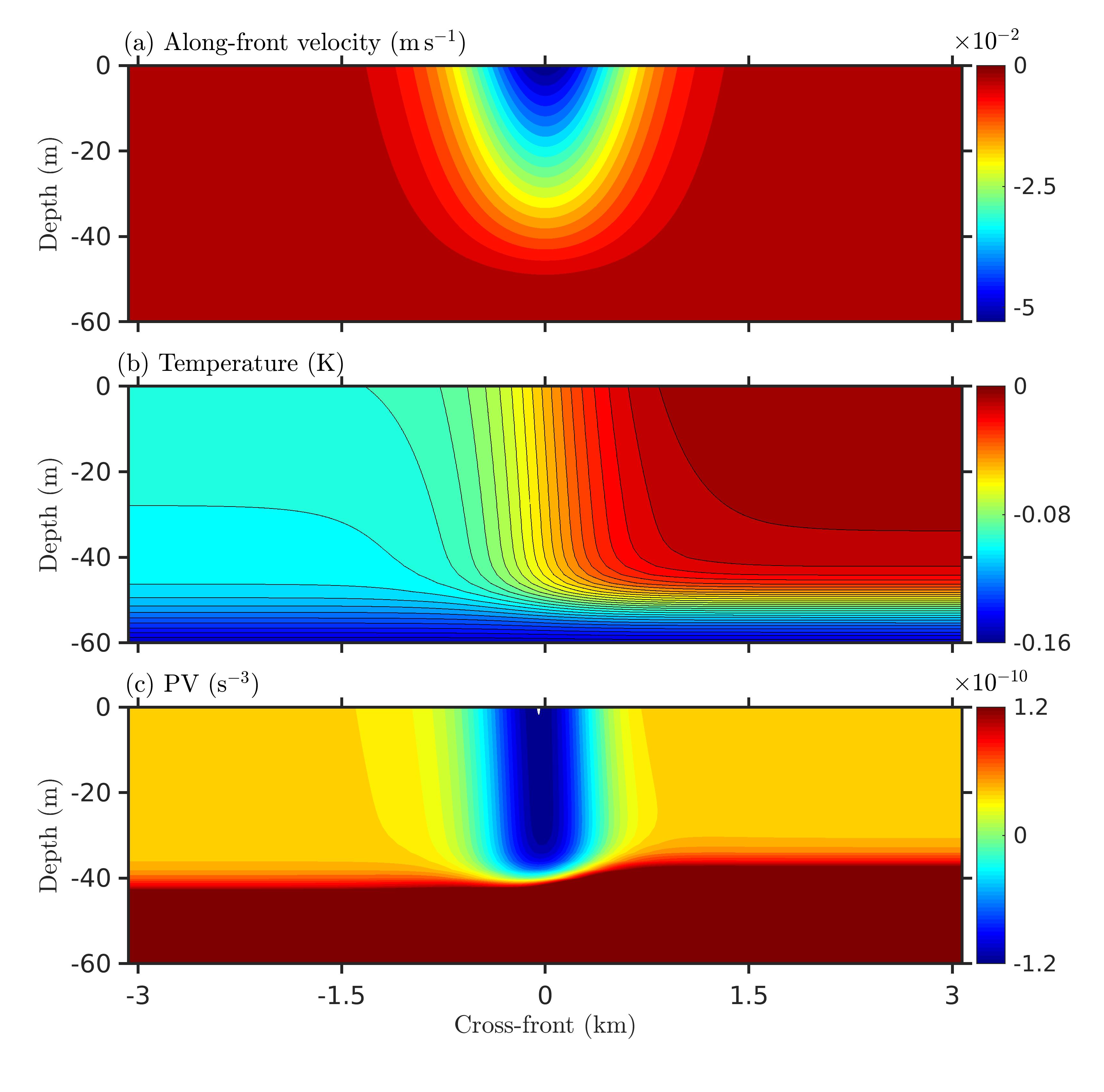}
 \caption{Initial profiles at the front: (a) along-front velocity,  (b) temperature and (c) potential vorticity.}
 \label{fig:initial_condition}
\end{figure}

\begin{figure}[!ht]
 \centering
 \includegraphics[width=\linewidth]{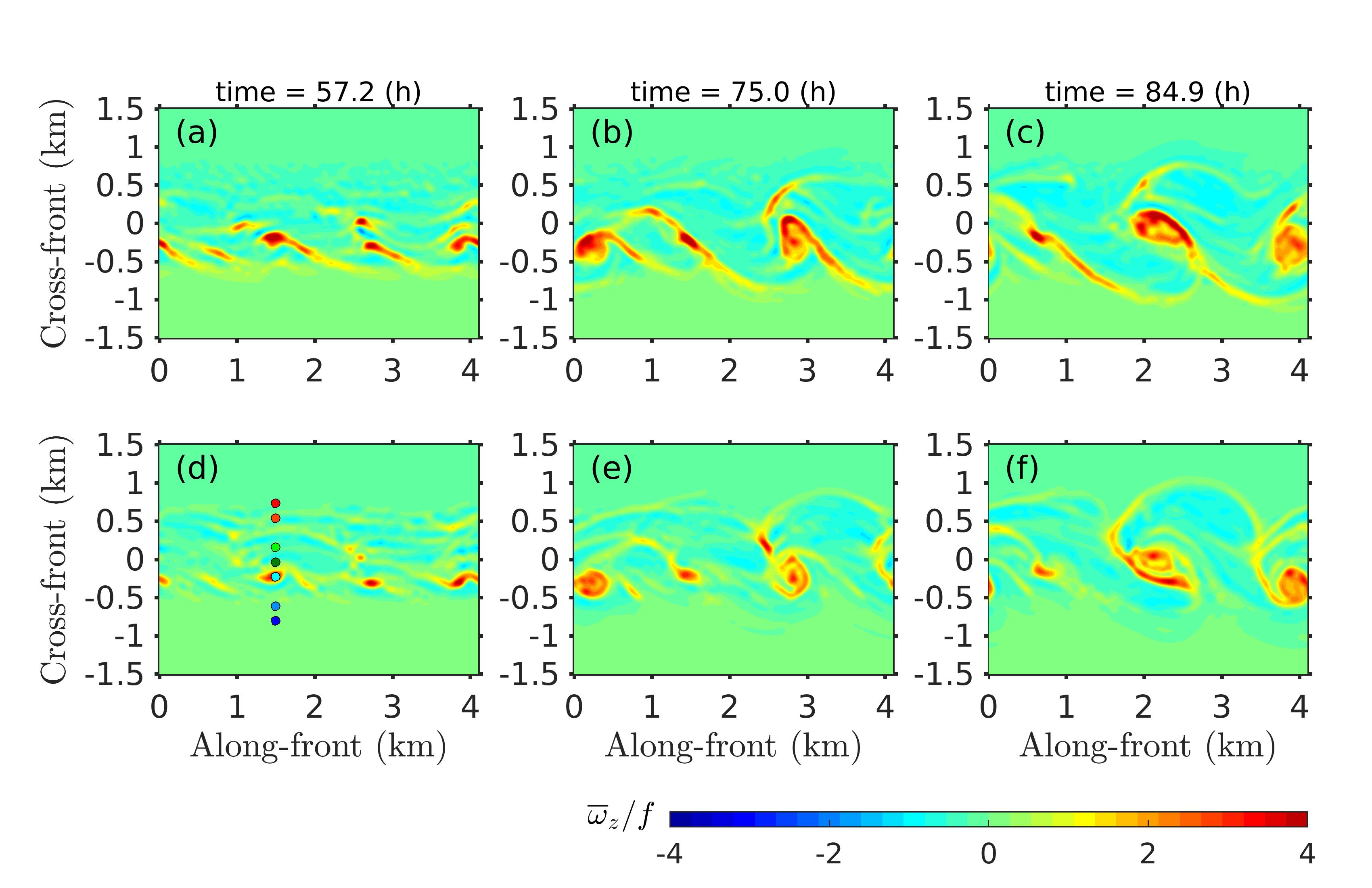}
 \caption{Evolution of coherent structures at the front. The figures show submesoscale vertical vorticity normalized by the Coriolis parameter at depths $10 \, \rm{m}$ (a, b, c) and $30 \, \rm{m}$ (d, e, f) at times $t = 57.2, 75$, and $84.9 \, \rm{h}$. In panel (d), solid circles depict the initial positions of the particles (P1-P7) whose trajectories are plotted in Fig.~\ref{fig:prt_trajectories}. Particles P1-P7 are arranged sequentially in the lateral with P1 at $y = -800 \, \rm{m}$.}
 \label{fig:omg3_depth_time}
\end{figure}

\begin{figure}[!ht]
 \centering
 \includegraphics[width=\linewidth]{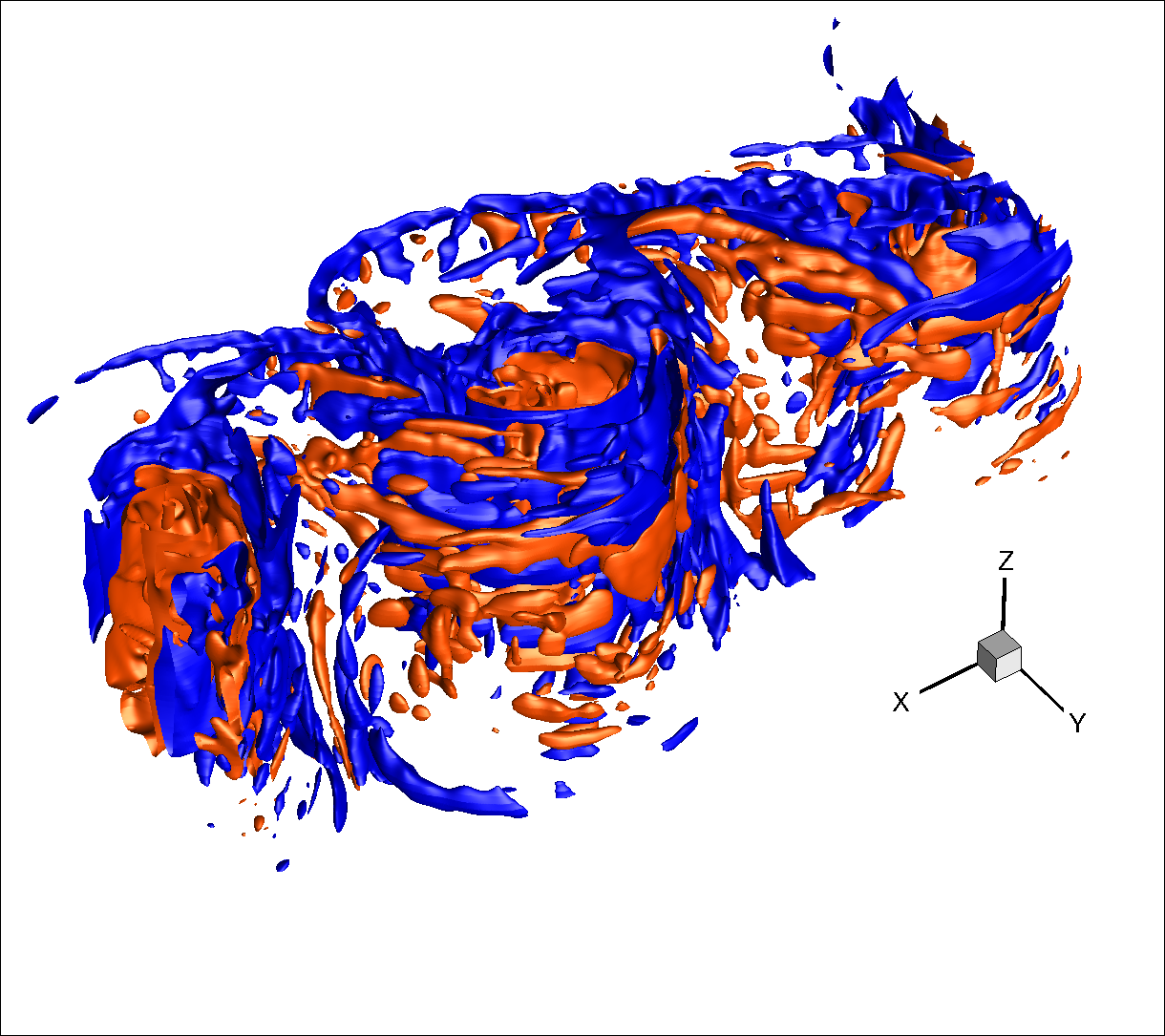}
 \caption{Visualization of coherent structures using the Q criterion on the submesoscale velocity field at $t = 84.9 \, \rm{h}$. The iso-surfaces of submesoscale Q are plotted at $\tilde{Q}/f^2 = 0.4$ (red) and $\tilde{Q}/f^2 = -0.4$ (blue).}
 \label{fig:coherent_structures_3d}
\end{figure}

\begin{figure}[!ht]
 \centering
 \includegraphics[width=\linewidth]{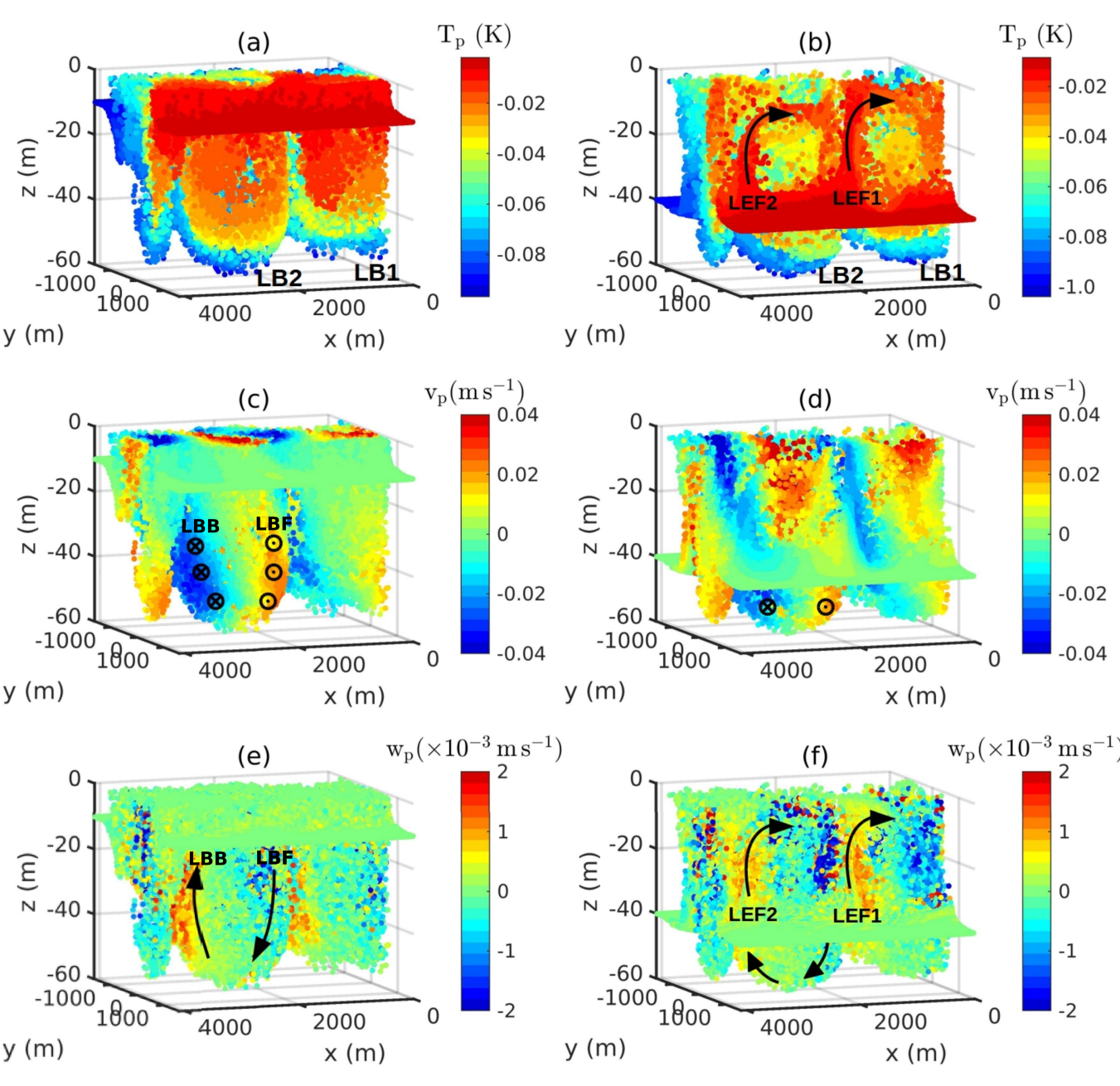}
 \caption{Plots of temperature (a, b), lateral velocity (c, d), and vertical velocity (e, f) at  $t = 84.9 \, \rm{h}$ corresponding to the particles released at $10 \, \rm{m}$ (left column) and $40 \, \rm{m}$ (right column) depth. In panels (a) and (b), LB1 and LB2 are the two particle lobes corresponding to the two eddies at the front. In panels (c) and (d), the symbols with dots inscribed within circles mark the side in LB2 where the lateral velocity of the particles is generally negative, whereas the symbols with crosses inscribed within the circles mark the side where the overall lateral velocity is negative. The overall upwelling/downwelling vertical velocity of the particles at the two sides of LB2 are depicted by arrows in panels (e) and (f). The arrows in panels (b) and (f), denoted as LEF1 and LEF2, identify the upwelling particle filaments.}   
 \label{fig:transport_structure_3D}
\end{figure}

\begin{figure}[!ht]
 \centering
 \includegraphics[width=\linewidth]{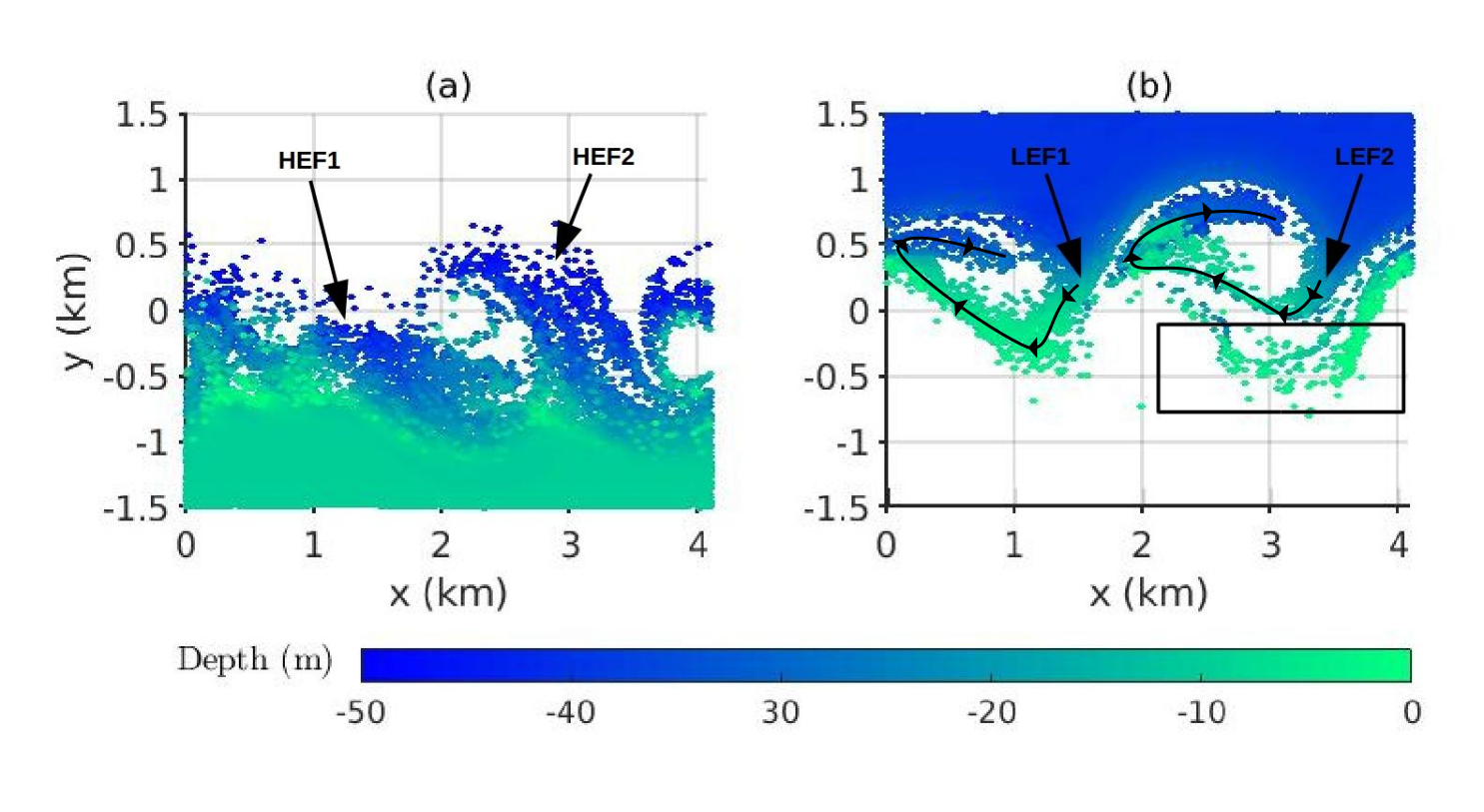}
 \caption{The $z$ coordinates after a flight time of 27.7 h is shown for particles released at $t = 57.2 \, \rm{h}$ and two different depths: (a) 10 m on  the heavy edge, $y <  -500$ m, and (b)  40 m on the light edge, $ y > 500$ m. In panel (a), HEF1 and HEF2 denote the downwelling of the heavy-edge particles mediated by filaments, and in panel (b), LEF1 and LEF2 denote the upwelling of the light-edge particles through filaments, also identified in Figs.~\ref{fig:transport_structure_3D}b,f. The solid black lines with arrows in panel (b) show the motion of the particles through filaments LEF1 and LEF2 with time, and the rectangular box encloses the particles which detach from the main branch LEF2 near the surface.}
 \label{fig:downwelling_upwelling}
\end{figure}

\begin{figure}[!ht]
 \centering
 \includegraphics[width=\linewidth]{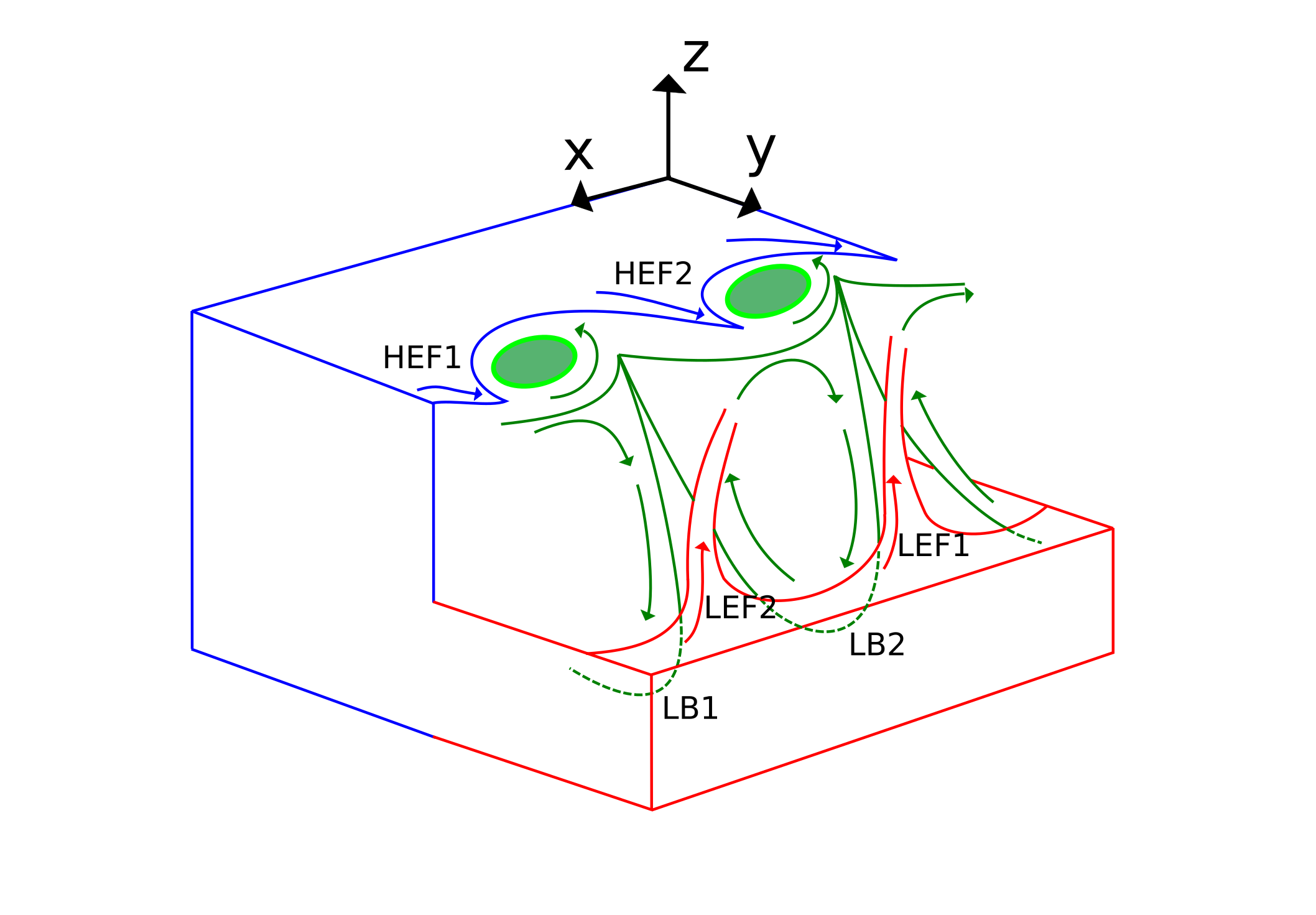}
 \caption{A schematic of the transport mediated by the coherent vortex filaments and eddies, and the circulation of particles organized within the lobes. The downward sloping regions of HEF1 and HEF2 are behind the lobes and hidden in this view.}
 \label{fig:transport_schematic}
\end{figure}

\begin{figure}[!ht]
 \centering
 \includegraphics[width=\linewidth]{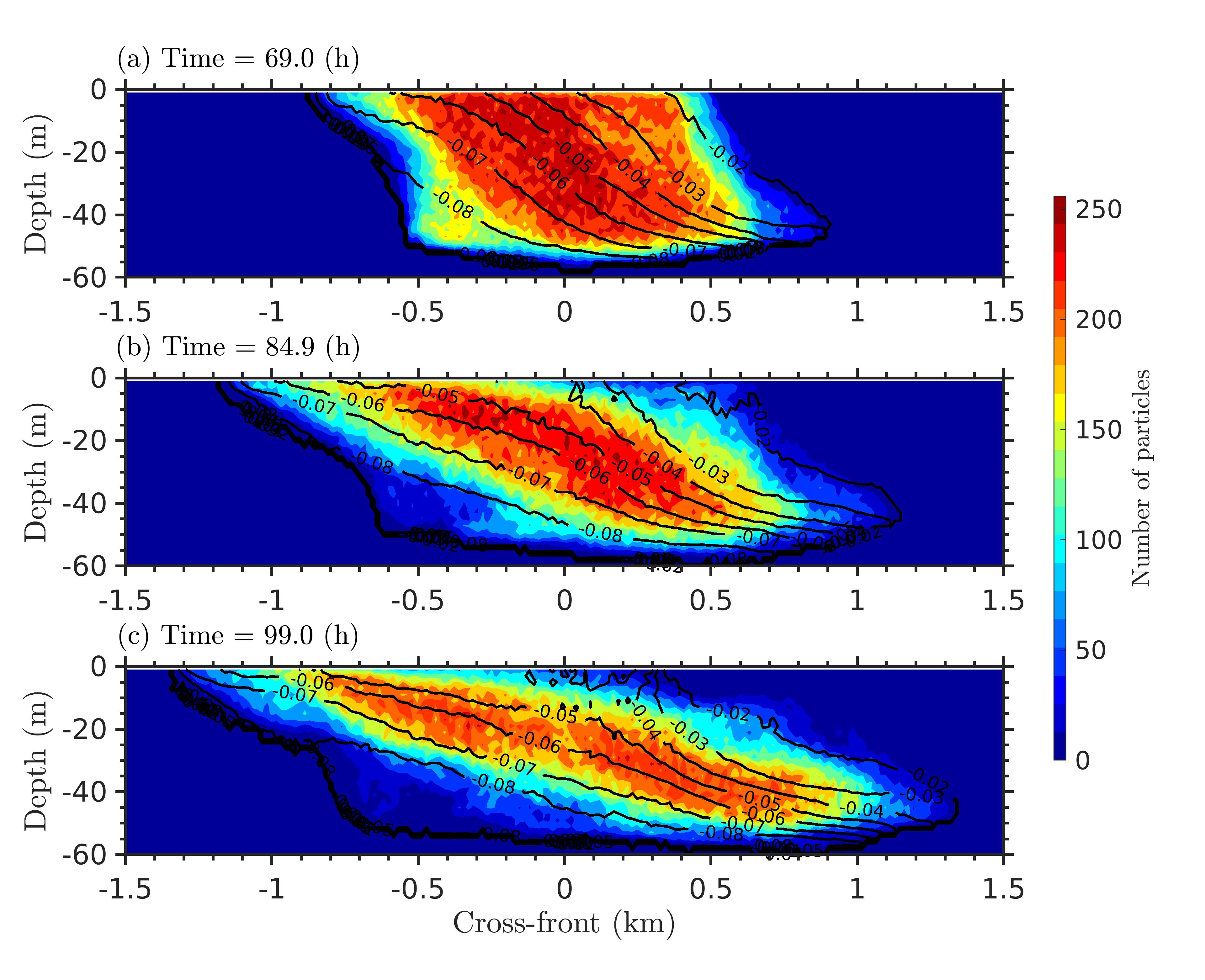}
 \caption{The organization of particles released in the central region with $-500 \, \rm{m} < y_p < 500 \, \rm{m}$ and $z_p > -50 \, \rm{m}$ at different times: (a) $t = 69 \, \rm{h}$,  (b) $84.9 \, \rm{h}$ and (c)  $99 \, \rm{h}$. The particles are sampled in the cells of a rectangular grid with the resolution $\Delta_s y = 16 \, \rm{m}$ in the lateral, and $\Delta_s z = 2 \,\rm{m}$ in the vertical. The solid lines in panels (a), (b) and (c) represent the isotherms corresponding to the mean temperature (along-front average) of the sampled particles.}
 \label{fig:front_adjustment}
\end{figure}

\begin{figure}[!ht]
 \centering
 \includegraphics[width=0.8\linewidth]{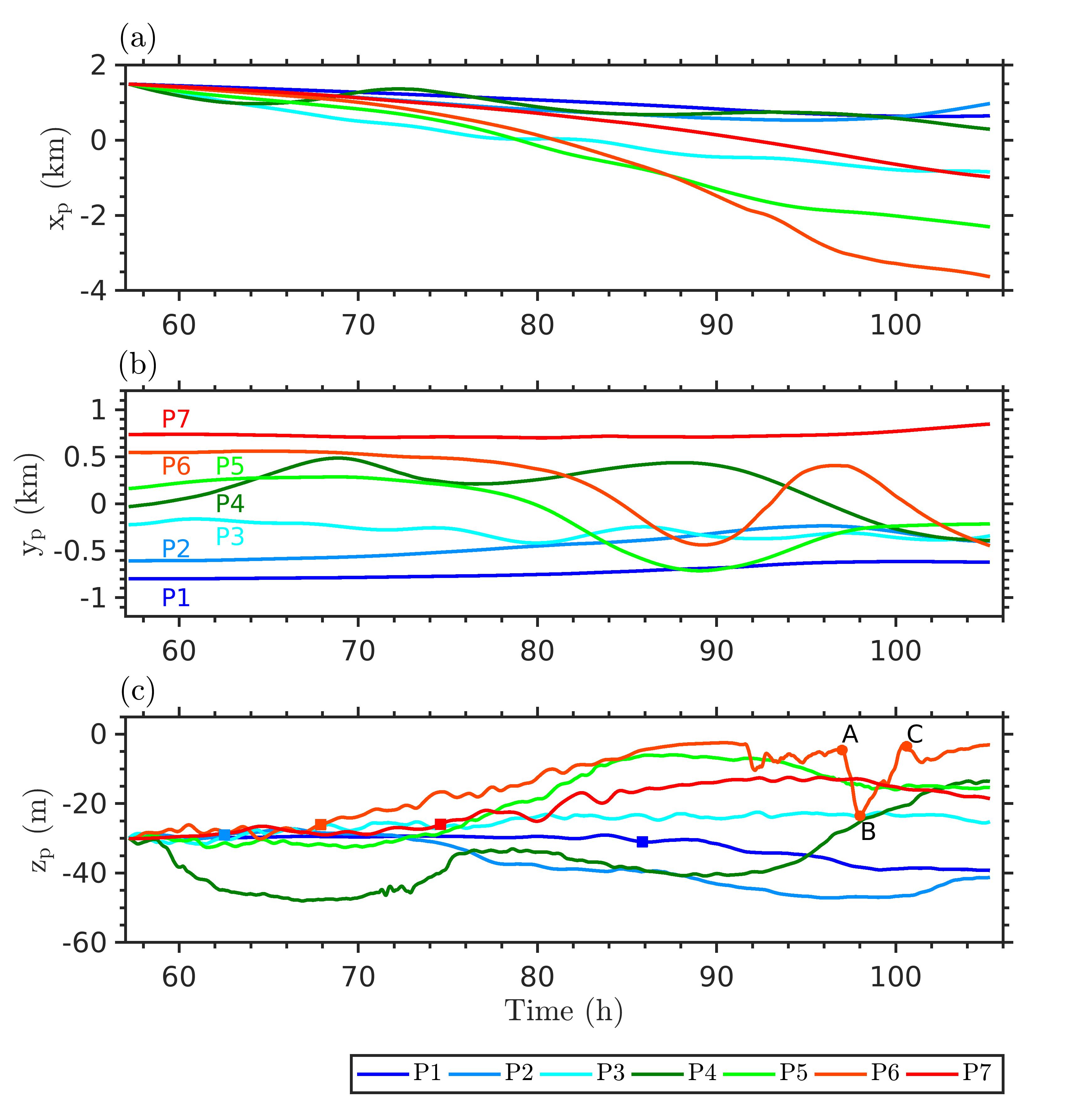}
 \caption{Trajectories  $(x_p(t), y_p(t), z_p(t))$ plotted in time for the particles released at a cross-front transect  (different $y$-locations)  through $x = 1490 \, \rm{m}$ and $z = -30 \, \rm{m}$. The initial positions of the particles in the $xy$-plane at $30 \, \rm{m}$ depth were shown in Fig.~\ref{fig:omg3_depth_time} (d). In panel (c), the solid squares in the vertical trajectories of edge-particles P1, P2, P6 and P7  denote the time when they start moving vertically. Points A, B, and C in the vertical trajectory of P6 mark the different phases of a rapid downwelling-upwelling event: A-B corresponds to the downwelling phase and B-C to the upwelling phase.}
 \label{fig:prt_trajectories}
\end{figure}

\begin{figure}[h!]
 \centering
 \includegraphics[width=0.6\linewidth]{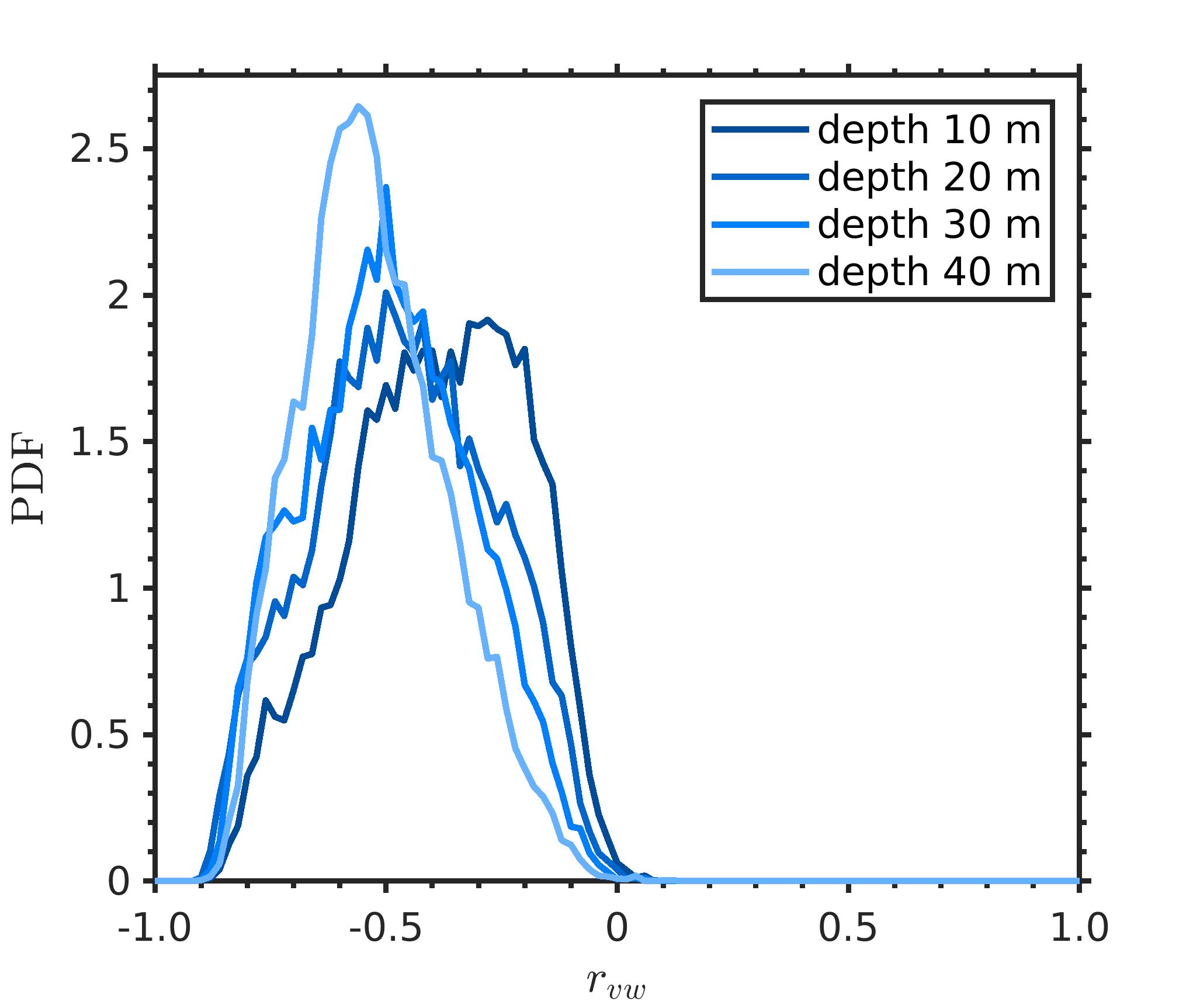}
 \caption{Probability density function (PDF) of the correlator $r_{xy}$ for  particles released in the central region at different depths:10 m, 20 m, 30 m, and 40 m.}
 \label{fig:pdf_r_vw}
\end{figure}

\begin{figure}[!ht]
 \centering
 \includegraphics[width=\linewidth]{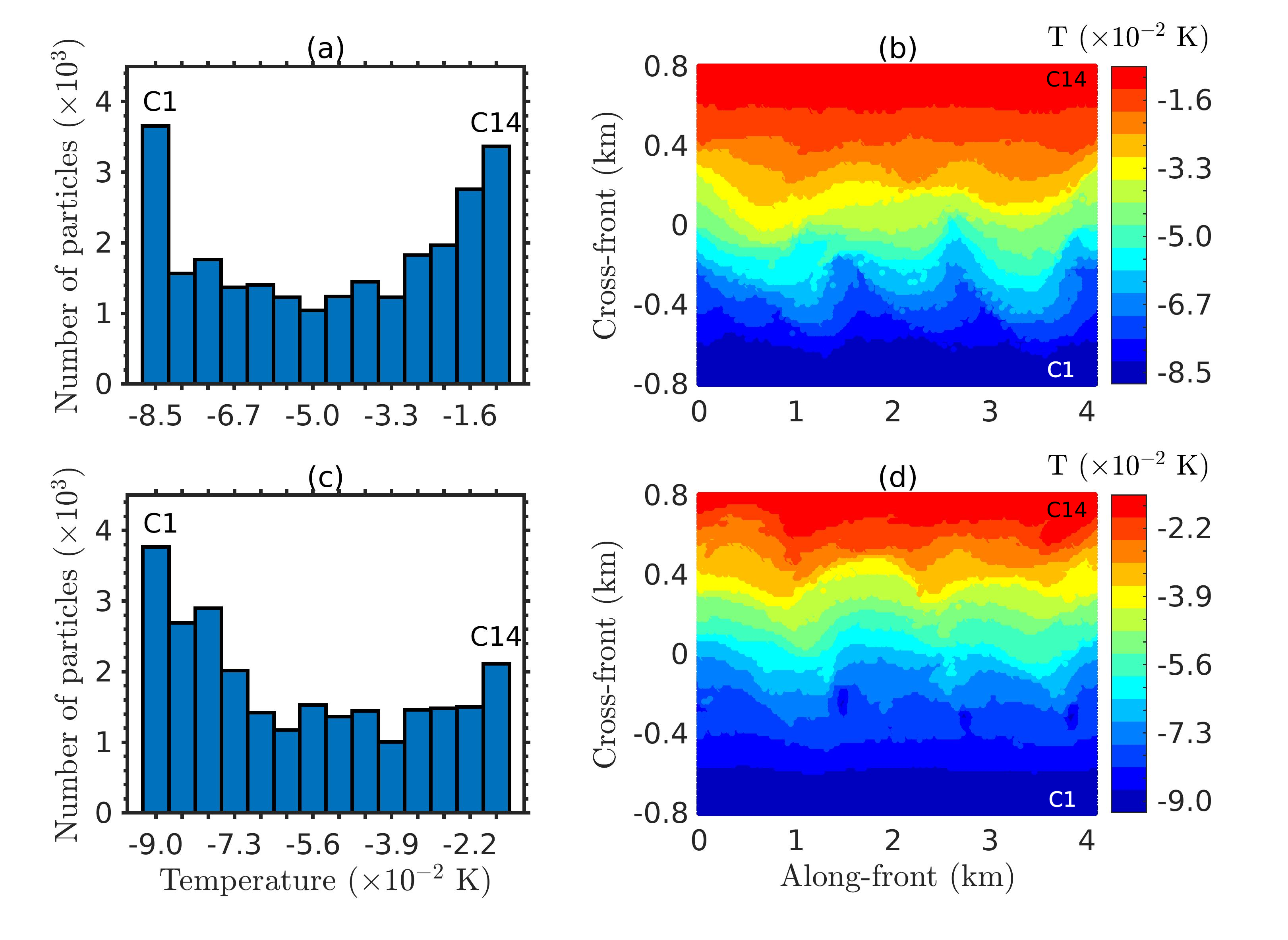}
 \caption{The initial configuration of the particle clouds released at $10\,\rm{m}$ and $40 \, \rm{m}$ depth: the mean temperature and the number of tracer particles in each cloud (a, c) and the organization of the clouds in the horizontal (b, d). Each particle cloud has particles with a similar density ranging from high (C1) to low (C14).  The particle clouds, especially in the central region, are in the form of long, thin meandering strips.}
 \label{fig:transport_setup}
\end{figure}

\begin{figure}[!ht]
 \centering
 \includegraphics[width=0.8\linewidth]{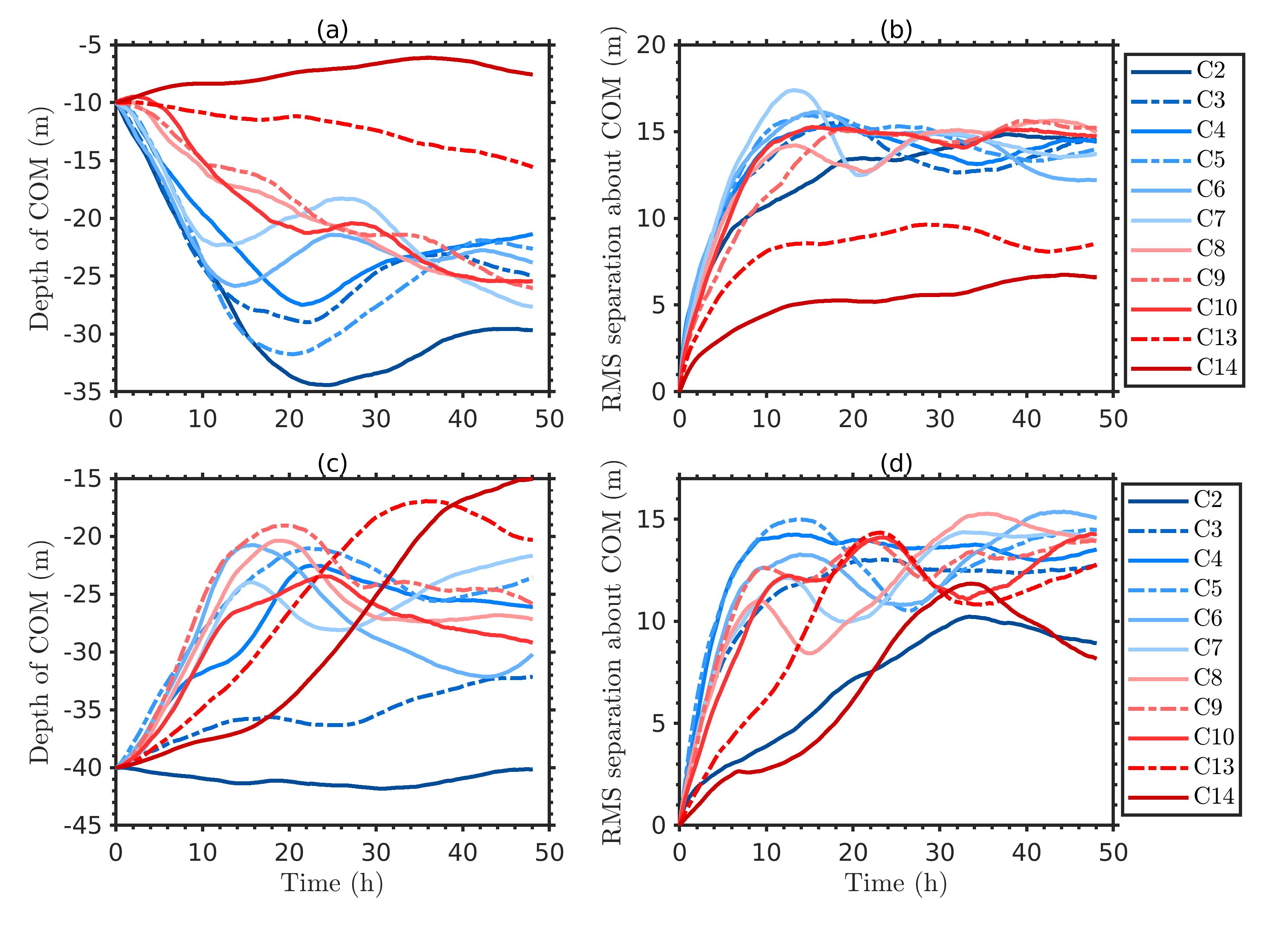}
 \caption{The vertical trajectories of the center of mass (COM) of the clouds released at $10 \, \rm{m}$ depth (a), and $40 \, \rm{m}$ depth (c). Also shown are the root-mean-square vertical displacements of constituent particles about the COM for the $10 \, \rm{m}$-depth release (b) and the $40 \, \rm{m}$-depth release (d).}
 \label{fig:vertical_transport}
\end{figure}

\begin{figure}[!ht]
 \centering
 \includegraphics[width=\linewidth]{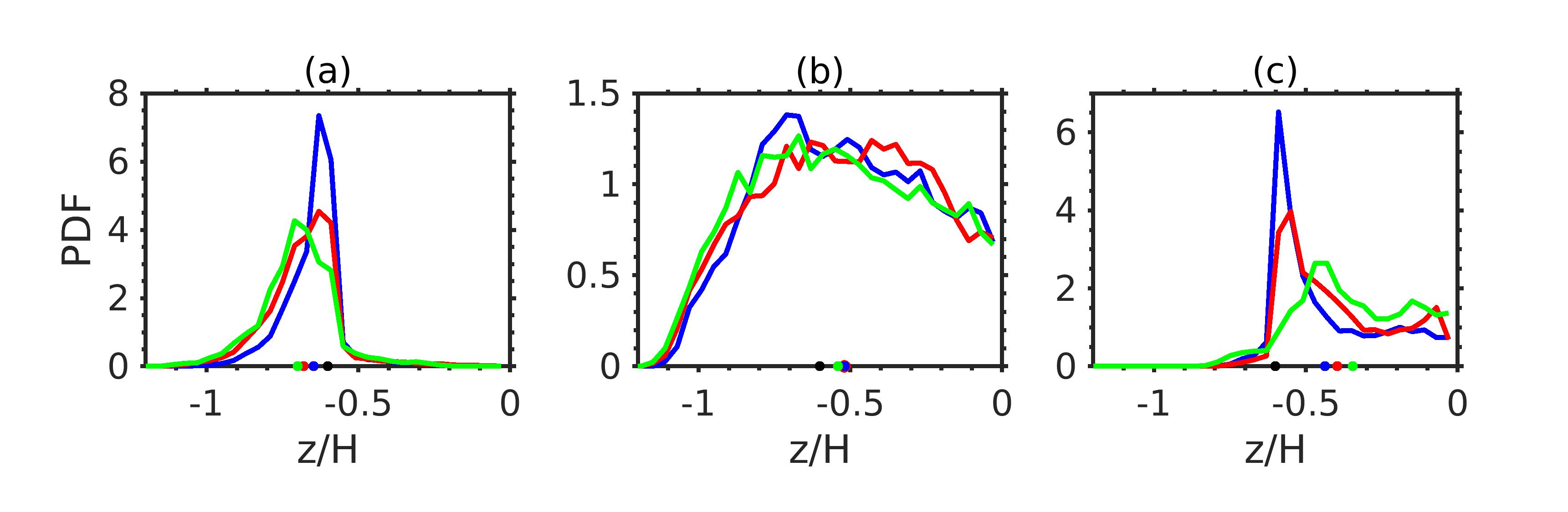}
 \caption{The probability density function (PDF) of the vertical distribution of the particles released at $30 \, \rm{m}$ depth  ($z/H = -0.6$) at: (a) the heavy edge, $-1000 \le y < -500 \, \rm{m}$, (b) the central region, $-500 \le y \le 500 \, \rm{m}$, and (c) the light edge, $500 < y \le 1000 \, \rm{m}$. The particles are released at $t = 57.2$ h and each panel shows the PDF at three different times: $t = 79.9 \, \rm{h}$ (blue line), $86.1 \, \rm{h}$ (green line) and $95 \, \rm{h}$ (red line). Solid colored circles on the horizontal axis depict the COM of the particles at the corresponding time and the black circles mark the initial COM.}
 \label{fig:vertical_distribution_pdf}
\end{figure}

\begin{figure}[!ht]
 \centering
 \includegraphics[width=\linewidth]{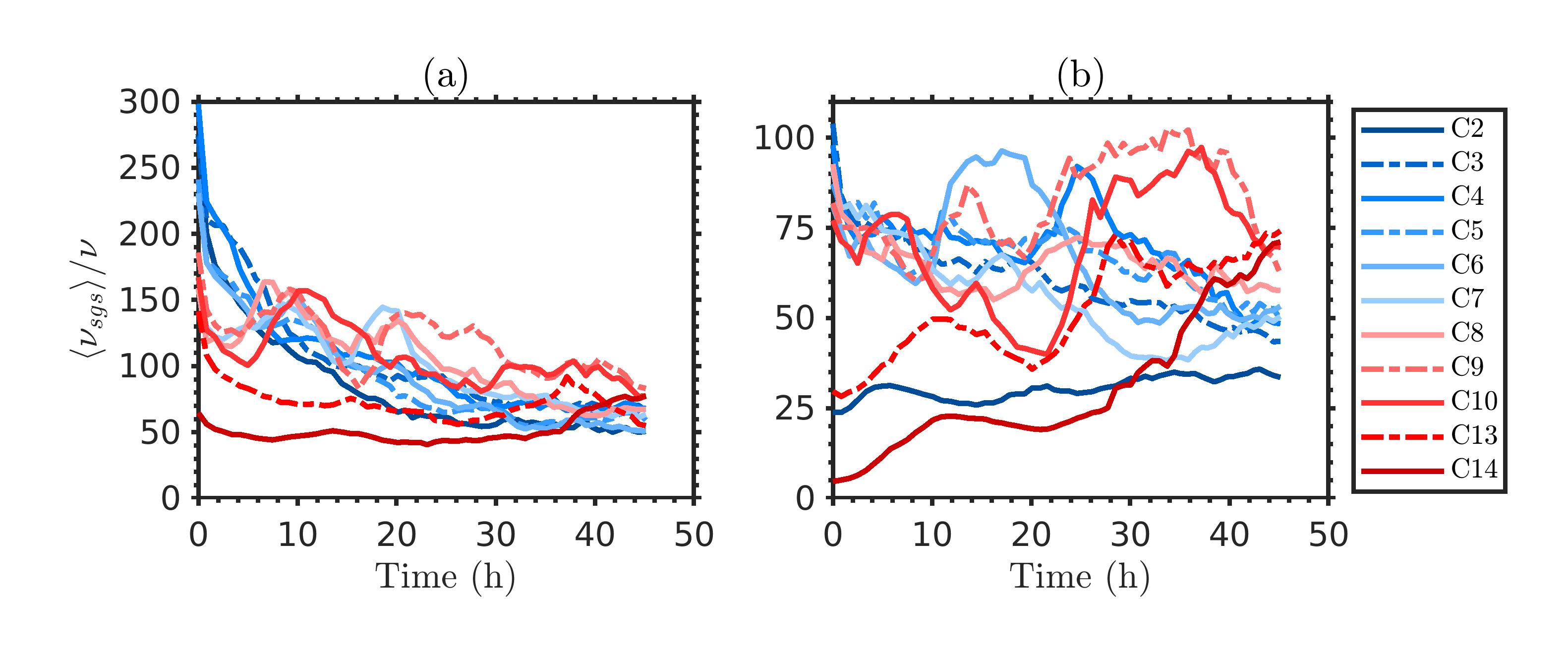}
 \caption{The mean subgrid viscosity experienced by the particle clouds released at (a) $10 \, \rm{m}$ and (b) $40 \, \rm{m}$ depth.}
 \label{fig:transport_nut}
\end{figure}

\begin{figure}[!ht]
 \centering
 \includegraphics[width=\linewidth]{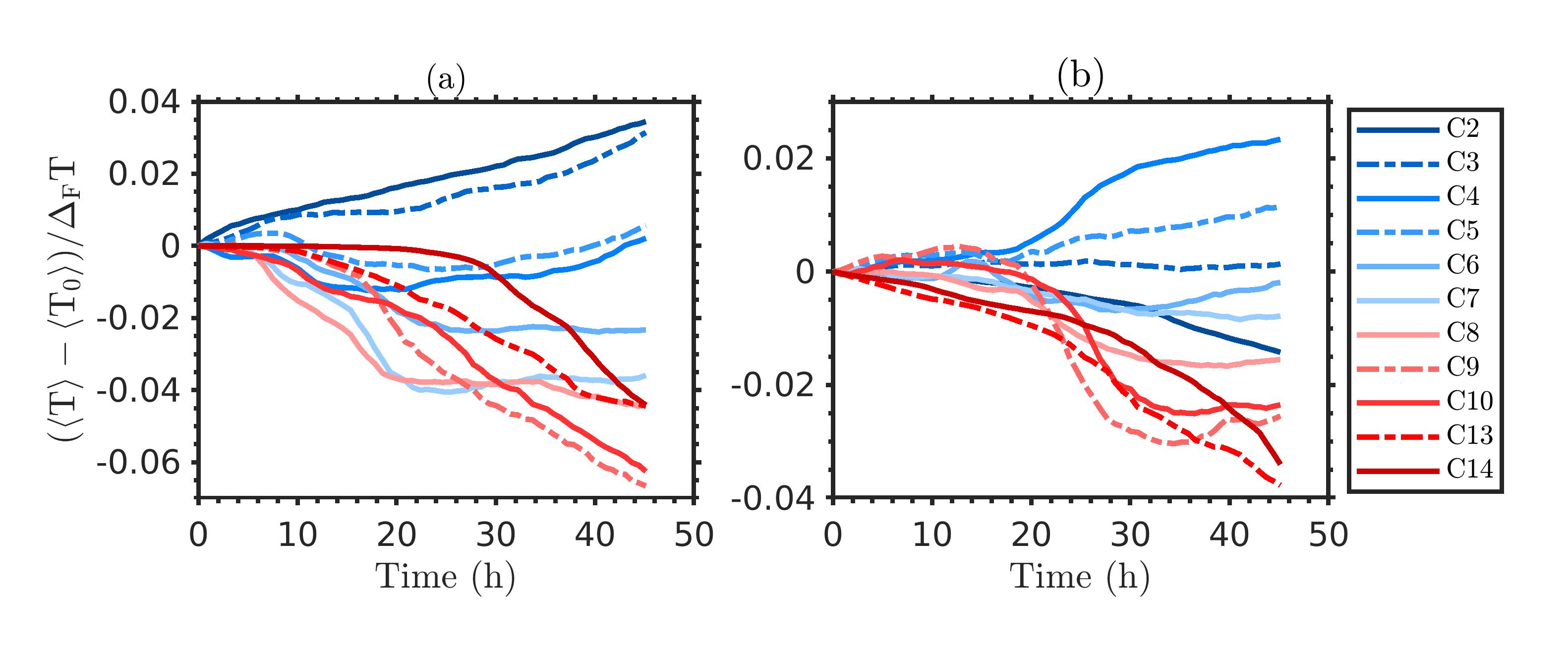}
 \caption{The change in mean temperature of the particle clouds released at (a) $10 \, \rm{m}$ and (b) $40 \, \rm{m}$ depth.}
 \label{fig:transport_rho}
\end{figure}

\begin{figure}[!ht]
 \centering
 \includegraphics[width=\linewidth]{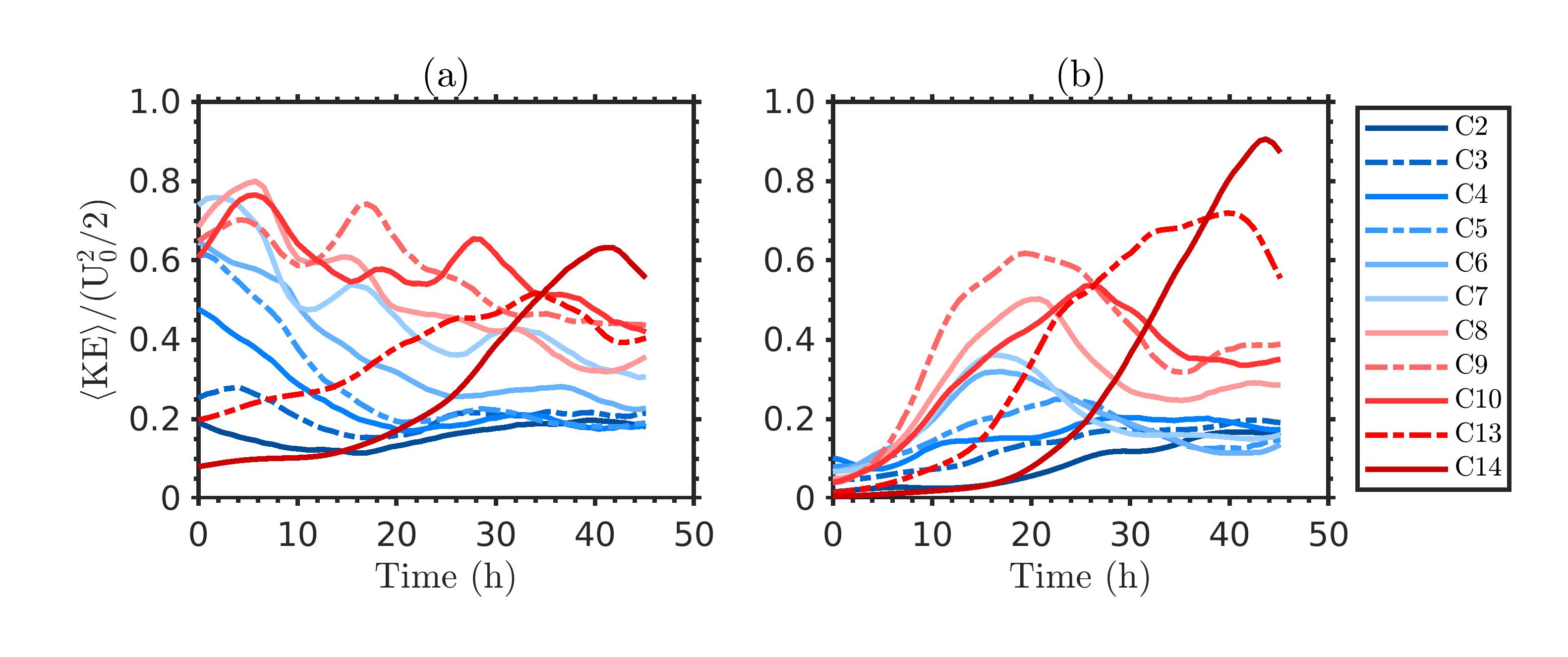}
 \caption{The mean kinetic energy of the particle clouds released at (a) $10 \, \rm{m}$ and (b) $40 \, \rm{m}$ depth.}
 \label{fig:transport_ke}
\end{figure}

\begin{figure}[!ht]
 \centering
 \includegraphics[width=0.6\linewidth]{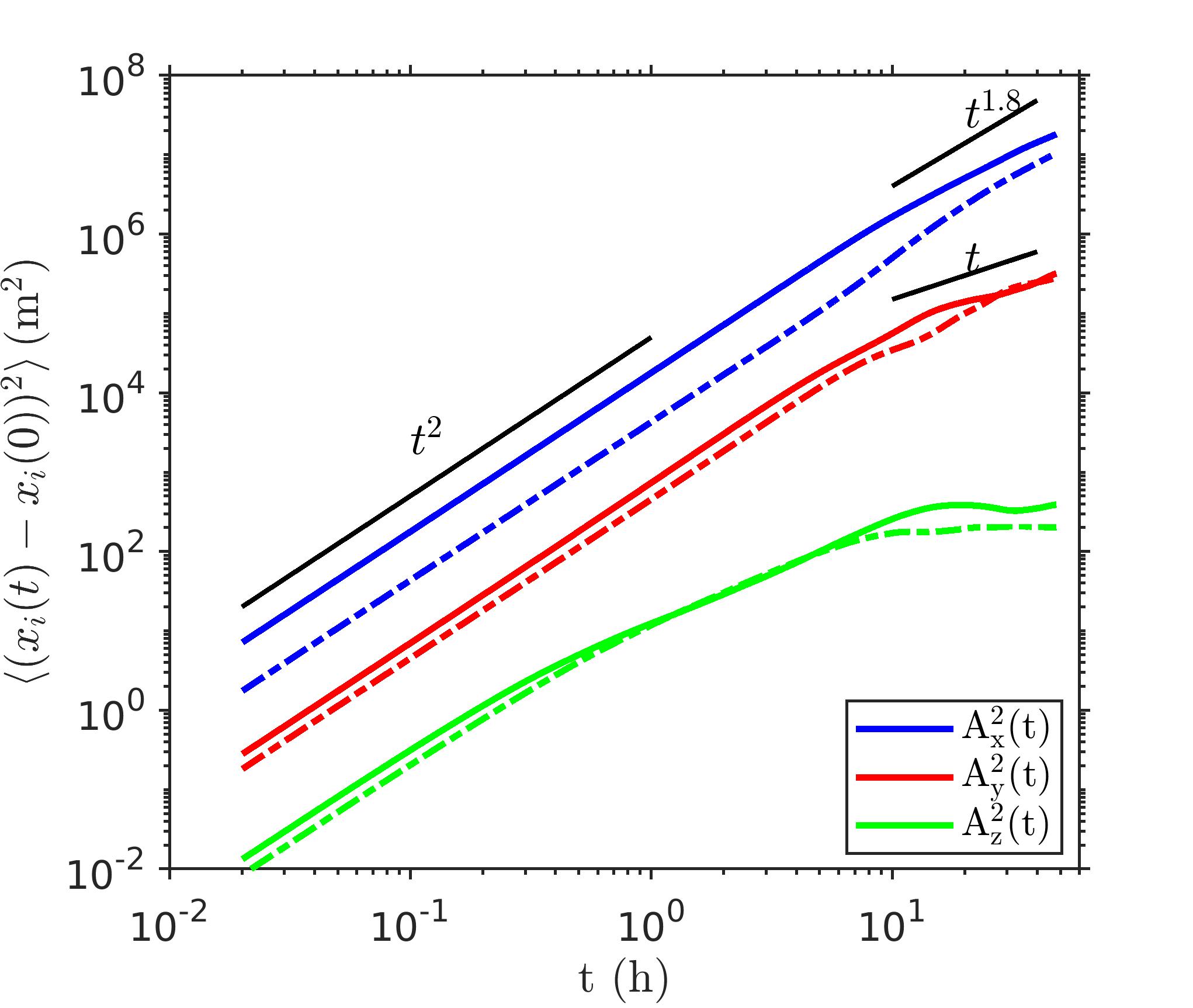}
 \caption{Absolute-dispersion components in $x$ (blue), $y$ (red) and $z$ directions (green) plotted as a function of time for the particles released in the central region, $-500 \, \rm{m} < y < 500 \, \rm{m}$, at 10 m (solid lines) and 30 m (dashed-dotted lines) depth.}
 \label{fig:absolute_dispersion}
\end{figure}

\begin{figure}[!ht]
 \centering
 \includegraphics[trim=1.0cm 0.2cm 0.2cm 1.0cm, width=\linewidth]{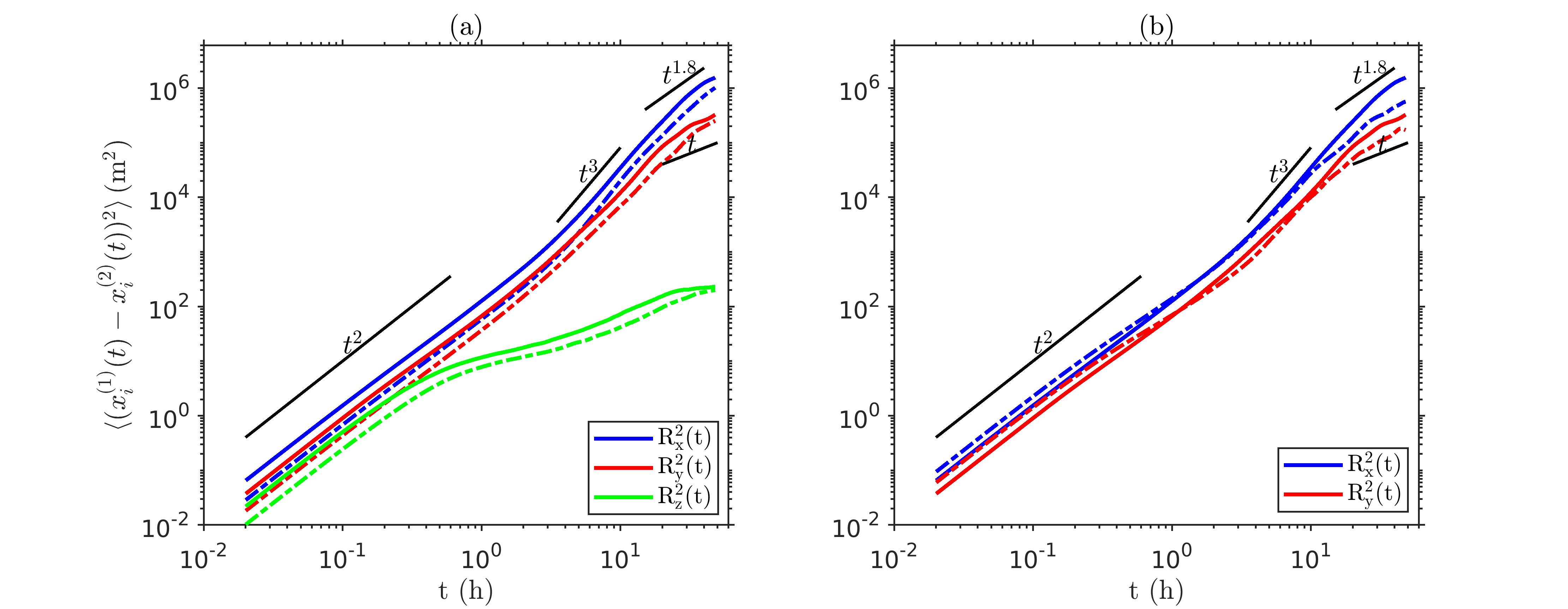}
 \caption{Relative-dispersion components in $x$ (blue lines), $y$ (red line) and $z$ directions (green lines) plotted as a function of time: (a) pairs released at  $10 \, \rm{m}$ (solid lines) and $30 \, \rm{m}$ (dashed-dotted lines) depth, and (b) pairs released at the surface (dashed-dotted lines) and $10 \, \rm{m}$ (solid lines) depth. For the surface particles, the $z$ component is zero because of the zero value of vertical velocity at the surface and is not plotted.
  In both (a) and (b), only those pairs released in the central region, $-500 \, \rm{m} < y < 500 \, \rm{m}$, are considered.} 
 \label{fig:relative_dispersion}
\end{figure}

\begin{figure}[!ht]
 \centering
 \includegraphics[trim=1.0cm 0.2cm 0.2cm 1.0cm, width=\linewidth]{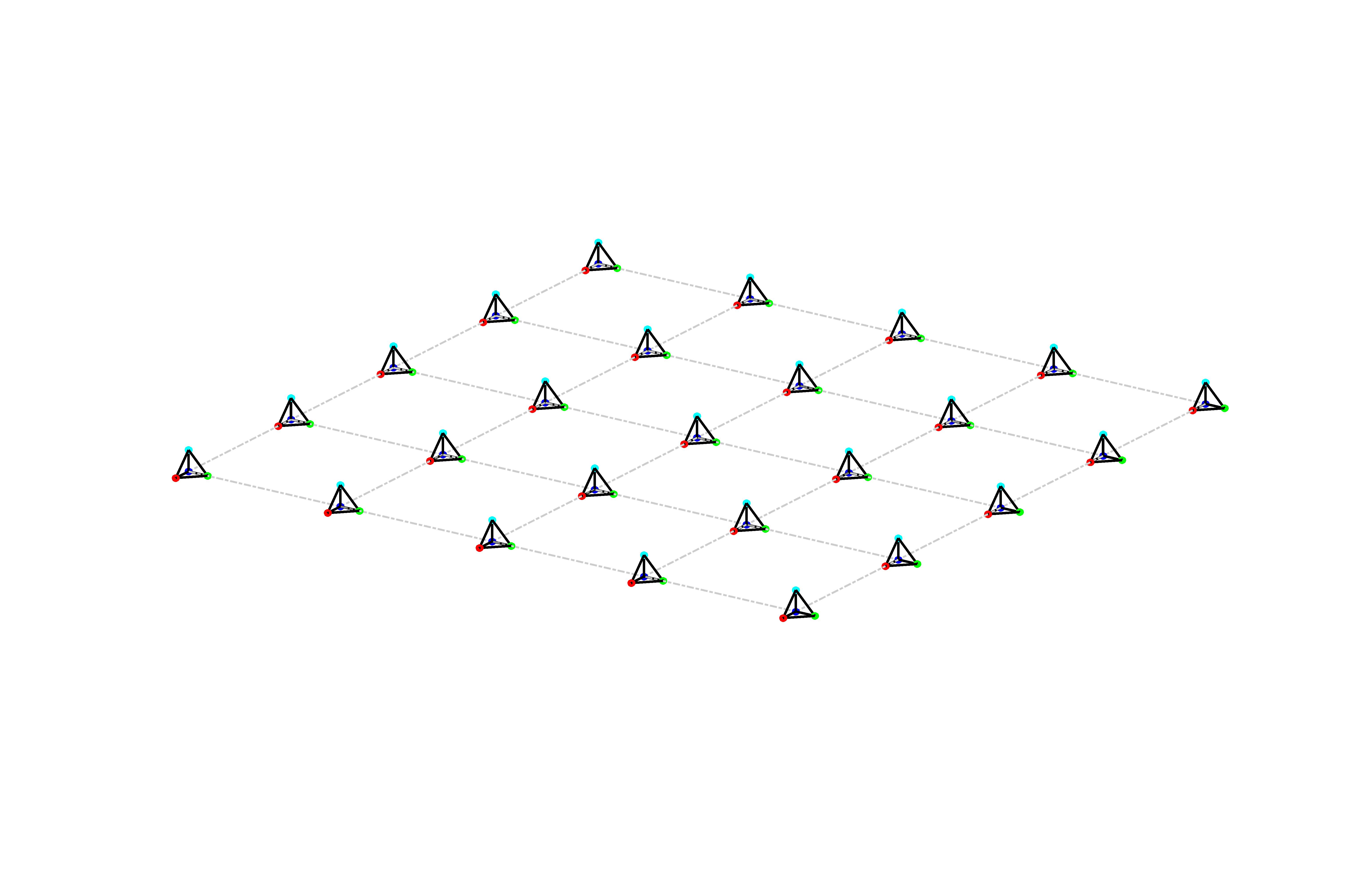}
 \caption{Construction of tetrads is illustrated by a small patch of the tetrads. Each tetrad is composed of four particles:  a node of the base-level particle lattice used for the single-particle statistics, a particle displaced by 2 m in the x direction, a  particle displaced by 2 m in the y direction, and a fourth particle from the lattice one level above, i.e., $2 \, \rm{m}$ above the base level.} 
 \label{fig:tetrad_construction}
\end{figure}

\begin{figure}[!ht]
 \centering
 \includegraphics[width=\linewidth]{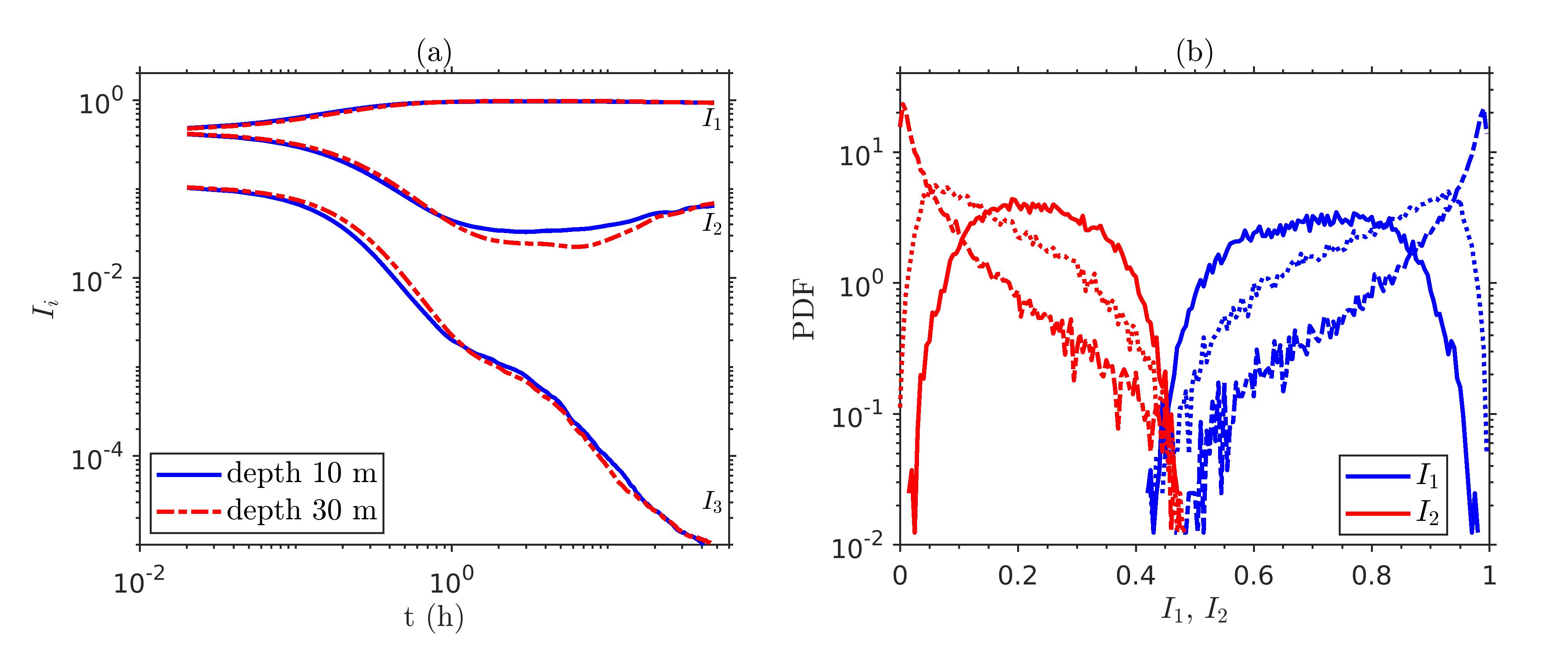}
 \caption{The distortion of the shape of tetrad particle clusters. (a) Normalized eigenvalues $I_1$, $I_2$ and $I_3$ as a function of time.  (b) PDFs of $I_1$ and  $I_2$ at different times after the release of the clusters: $\Delta t = 20\,\rm{min}$, solid lines; $\Delta t = 40\,\rm{min}$, dotted lines; $\Delta t = 1.2\,\rm{h}$, dotted-dashed lines.}
 \label{fig:multiparticle_dispersion}
\end{figure}

\end{document}